\documentclass[12pt,superscriptaddress,floatfix,preprintnumbers,nofootinbib]{revtex4}              
\pdfoutput=1
\usepackage{graphicx}
\usepackage{amssymb}
\usepackage{rotating}
\usepackage{epsfig}
\usepackage[lofdepth,lotdepth,caption=false]{subfig}
\usepackage{color}

\begin{document}

\title{Probing Non-standard Interaction of Neutrinos with IceCube and DeepCore}
\author{Arman~Esmaili} 
\affiliation{Instituto de Fisica Gleb Wataghin - UNICAMP, 13083-859, Campinas, SP, Brazil}
\author{Alexei~Yu.~Smirnov} 
\affiliation{Abdus Salam International Centre for Theoretical Physics, ICTP, I-34010, Trieste, Italy}

\begin{abstract}

We consider effects of the Non-Standard Interactions (NSI)  on oscillations of the high energy atmospheric neutrinos. The $\nu_\mu-$oscillograms are constructed and their dependence on the NSI strength parameters $\epsilon_{\alpha \beta}$ studied. We computed the zenith angle distributions of the $\nu_\mu-$events in the presence of NSI in different energy regions. The distributions are confronted with  the IceCube-79 (high energy sample) and the DeepCore (low energy sample) data and constraints on the strength  parameters $|\epsilon_{\mu \tau}| \lesssim 6 \times 10^{-3}$ and $|\epsilon_{\mu\mu}-\epsilon_{\tau\tau}| \lesssim 3 \times 10^{-2}$ (90\% C.L.)  have been obtained. Future measurements of the zenith angle distributions by DeepCore in several energy bins will allow to improve the bounds by factor 2 - 3. We discuss the signatures of NSI in the zenith angle and energy distributions of events which allow to discriminate them from the effects of sterile neutrinos.

\end{abstract}

\date{\today}

\maketitle

\section{Introduction}
\label{sec:intro}

Searches for effects of Non-Standard Interactions (NSI) of neutrinos is one of the avenues to uncover new physics beyond the standard model (see \cite{GonzalezGarcia:2004wg,Ohlsson:2012kf} for review). The NSI can show up in neutrino production and absorption or scattering in a source and detector. They can affect neutrino propagation in medium modifying the standard oscillation pattern. The NSI effects in production (absorption) and in propagation may, or may not,  be related and interplay of these effects is possible.

At the phenomenological level the NSI of neutrinos are described by the strength parameters, $\epsilon_{\alpha \beta}$, which enter the effective potentials $\epsilon_{\alpha \beta} r V_{CC}$ generated by the coherent $\nu_{\alpha} \rightarrow \nu_\beta$ forward scattering. Here $V_{CC} \equiv \sqrt{2}G_F n_e$, $G_F$ is the Fermi coupling constant and $r\equiv n_d/n_e$, where $n_d$ and $n_e$ are the number densities of $d$-quarks and electrons correspondingly. 

Effects of NSI in the atmospheric neutrinos have been considered in a number of publications before (see~\cite{atmnsi} and references therein). Also, the effect of NSI on propagation of accelerators, solar and supernova neutrinos have been studied (see~\cite{Ohlsson:2012kf} and references therein). Furthermore, using various restrictions~\cite{bounds} derived from oscillation data, neutrino scattering off charged leptons and quarks, lepton flavor violating processes and electron-positron scattering the following model-independent bounds on the NSI strength parameters have been obtained in~\cite{Biggio:2009nt}:
\begin{equation}
\epsilon_{\mu \tau} < 0.21, ~~~
\epsilon_{\mu \mu} < 0.046, ~~~
\epsilon_{\tau \tau} < 9. 
\end{equation}

The Super-Kamiokande collaboration performed a dedicated experimental search of NSI in the atmospheric neutrinos crossing the Earth~\cite{Mitsuka:2011ty}. No effect has been found which leads to the upper bounds (at 90\% C.L.)
\begin{equation}
|\epsilon_{\mu \tau}| < 1.1 \times 10^{-2},~~~~
|\epsilon^{\prime}| = |\epsilon_{\mu \mu} - \epsilon_{\tau \tau}| < 4.9 \times 10^{-2}.
\label{eq:sk-limit}
\end{equation}
Recently, using both neutrino and antineutrino beams, the MINOS collaboration has obtained bound on the off-diagonal parameter $-0.20 < \epsilon_{\mu\tau} < 0.07$ ($90\%$ C.L.)~\cite{Adamson:2013ovz}.  

In~\cite{Ohlsson:2013epa}, the possibility to constrain NSI in the proposed PINGU extension of the IceCube detector has been explored. It was shown that, due to the high efficiency in detection of low energy ($1-20$~GeV) atmospheric neutrinos, PINGU may have good sensitivity to $\epsilon_{\mu\tau}$.
   
It is expected that NSI produce sub-leading effects in flavor oscillations of neutrinos. However, in certain situations their effects can dominate or be enhanced. It is well-known that the standard oscillations in matter disappear with increase of neutrino energy. Indeed, for the $\nu_e-$oscillation modes, the mixing in matter decreases with energy as $\sin^2 2\theta_m \propto (2E_\nu V_{CC})^{-2}$, although the oscillation length converges to the refraction length which is comparable to the Earth's radius. In contrast, for the $\nu_\mu-$ and $\nu_\tau-$ modes at high energies the mixing angle is unsuppressed. However, in this case the oscillation length (which nearly coincides with the vacuum oscillation length) increases with energy and becomes much bigger than the size of the Earth. Consequently, the oscillation phase and therefore the oscillation effects become very small.

In the presence of non-standard interactions the situation can be different, if the off-diagonal elements are nonzero. In this case oscillations do not disappear with increase of the neutrino energy, and their characteristics will be determined by the matter part of Hamiltonian. This means that appearance of oscillation effects at high energies will testify for NSI. At very high energies (which are inversely proportional to $\epsilon$) NSI dominate over standard oscillations and in this range their signal is simple and clear. On the other hand, in the range $E_\nu \sim (20 - 100)$ GeV, NSI can interfere with standard oscillations which leads to enhancement of the NSI effects. 

In this connection we propose to search for NSI at energies $E_\nu > 20$~GeV with huge atmospheric neutrino detectors: the IceCube and DeepCore. We show that the already collected statistics in these experiments allows to substantially improve the current bounds in Eq.~(\ref{eq:sk-limit}). 

The paper is organized as follows. In Sec.~\ref{sec:prob} we compute and study properties of the oscillation probabilities in the presence of NSI. We construct the corresponding oscillograms and explore their dependence on the NSI strength parameters. In Sec.~\ref{sec:cons} we compute the zenith angle distributions of the $\nu_\mu-$events in the low and high energy regions. We use the IceCube-79 and DeepCore data to obtain bounds on the NSI strength parameters. Future sensitivity of DeepCore to these parameters is evaluated. In Sec.~\ref{sec:discriminate} we discuss how the NSI effects can be disentangled from the possible effects of sterile neutrinos. Conclusions are presented in Sec.\ref{sec:conc}. 

\section{Oscillation probabilities in the presence of NSI}
\label{sec:prob}

\subsection{NSI strength parameters}

NSI's modify neutrino forward scattering, and consequently, the usual pattern of flavor oscillations. In the presence of NSI evolution of neutrinos in matter can be described by the Hamiltonian
\begin{equation}
\label{eq:hamiltonian1}
H_{3\nu} = \frac{1}{2E_\nu} 
U_{\rm PMNS} M^2U_{\rm PMNS}^\dagger +{\rm diag}(V_{CC},0,0) + \sum_f V_f\epsilon^f~,
\end{equation} 
where $U_{\rm PMNS}$ is the PMNS mixing matrix, $M^2 \equiv {\rm diag}\{0,\Delta m^2_{21},\Delta m^2_{31}\}$, and $\Delta m^2_{ij} \equiv  m_i^2 - m_j^2$ are the neutrino mass-squared differences. The last term of Eq.~(\ref{eq:hamiltonian1}) is the matter potential resulting from the NSI of neutrinos. The contribution from neutrino scattering on fermion of type $f$,  $\nu_\alpha+f\to\nu_\beta+f$, is given by $V_f\epsilon^f$, with $V_f \equiv \sqrt{2}G_Fn_f$, where $n_f$ is the number density of fermion $f$ and $\epsilon^f$ is the matrix of NSI strength parameters. For antineutrinos the sign of potentials would change, $V \rightarrow -V$, and mixing matrix equals the conjugate one:~$U_{\rm PMNS}^\ast$.

Normalizing the density of fermions $n_f$ by the density of $d-$quarks, $n_d$, we define the total strength for a given medium as 
\begin{equation}
\epsilon\equiv\sum_f \frac{n_f}{n_d}\epsilon^f~.
\end{equation}
Then the NSI term in Eq.~(\ref{eq:hamiltonian1}) can be rewritten as $V_d \epsilon = r V_{CC} \epsilon$, where $r\equiv n_d/n_e$. For the Earth we have $n_n \approx n_p$  (departure from equality is less than 3\%) and therefore $r=3$. Finally, the Hamiltonian takes the form:
\begin{equation}
\label{eq:hamiltonian2}
H_{3\nu} = \frac{1}{2E_\nu} U_{\rm PMNS}M^2U_{\rm PMNS}^\dagger + 
V_{CC} {\rm diag}(1, 0, 0) + rV_{CC}\epsilon~, 
\end{equation} 
where the Hermitian matrix of NSI strength parameters $\epsilon$ can be written as 
\begin{equation}
\epsilon=
\left(
\begin{array}{ccc}
\epsilon_{ee}  & \epsilon_{e\mu}  & \epsilon_{e\tau}  \\
 \epsilon_{e\mu}^\ast &  \epsilon_{\mu\mu} & \epsilon_{\mu\tau}  \\
 \epsilon_{e\tau}^\ast & \epsilon_{\mu\tau}^\ast  & \epsilon_{\tau\tau}  
\end{array}
\right)~. 
\end{equation}
The hermiticity of $\epsilon$ implies that the diagonal elements of the matrix are real and we assume that the non-diagonal elements are also real: $\epsilon_{\alpha\beta}=\epsilon_{\alpha\beta}^\ast$. Furthermore, for simplicity we will neglect the elements of the first raw and column, assuming that $\epsilon_{e\beta} \ll \epsilon_{\mu \tau}$ for $(\beta = e, \mu, \tau)$. 

The oscillation probabilities $P(\nu_\mu\to\nu_\mu)$ and $P(\nu_\mu\to\nu_\tau)$, which we are mainly interested in this paper, are sensitive to $\epsilon_{\mu\tau}$, that quantifies strength of the flavor changing neutral current interaction $\nu_\mu+f\to\nu_\tau+f$, and $\epsilon^\prime \equiv \epsilon_{\tau\tau}-\epsilon_{\mu\mu}$, which gives the non-universality of $\nu_\mu$ and $\nu_\tau$ neutral current interactions. To find the oscillation probabilities in the presence of NSI, we solve numerically the $3\nu$ evolution equation with the Hamiltonian in Eq.~(\ref{eq:hamiltonian2}). For the density profile of the Earth we use the PREM model~\cite{prem}. We take the best-fit values of oscillation parameters from the global fit in~\cite{GonzalezGarcia:2012sz}: 
$\sin^2\theta_{13}=0.023$, $\sin^2\theta_{12}=0.30$, $\sin^2\theta_{23}=0.4$, 
$\Delta m^2_{21}=7.5\times10^{-5}~{\rm eV}^2$ and $\Delta m^2_{31}=2.4\times10^{-3}~{\rm eV}^2$. 

In general, the parameters $\epsilon_{\alpha\beta}$ depend on composition of medium and so change on the way of neutrinos. We will neglect this dependence.

\subsection{Features of oscillations in $2\nu$ approximation}

If $\epsilon_{ee}$, $\epsilon_{e\mu}$, $\epsilon_{e\tau} \ll 1$, then at energies much above the 1-3 resonance ($E_\nu > 20$~GeV) the state $\nu_{3m} \approx \nu_e$ ($\bar{\nu}_{1m} \approx \bar{\nu}_e$) decouples from the evolution of the rest of the $3\nu$ system. In this case results from the $2\nu$ system provide good approximation and allow us to understand the main features of NSI effects. 

In the $2\nu$ case the effective Hamiltonian can be written as 
\begin{equation}
\label{eq:h2nu}
H_{2\nu}=\frac{\Delta m_{31}^2}{2E_\nu}U(\theta_{23})  
\left(
\begin{array}{cc}
 0 & 0  \\
  0 & 1 
\end{array} \right) U^\dagger (\theta_{23}) 
+V_d \left(
\begin{array}{cc}
 \epsilon_{\mu\mu} & \epsilon_{\mu\tau}  \\
 \epsilon_{\mu\tau} & \epsilon_{\tau\tau}  
\end{array}
\right)~,
\end{equation}
where $U(\theta_{23})$ is the $2\times2$ rotation matrix with the angle $\theta_{23}$. At very high energies, ${\Delta m_{31}^2}/{2E_\nu} \ll~V_d \epsilon$, the second term dominates and dynamics of propagation is completely determined by matter effects. In this limit diagonalization of the Hamiltonian in Eq.~(\ref{eq:h2nu}) gives the mixing angle  
\begin{equation}
\label{eq:parameters}
\sin2\xi = \frac{2\epsilon_{\mu\tau}}{\sqrt{4\epsilon_{\mu\tau}^2+
\epsilon^{\prime2}}}~, 
\end{equation}
and the level splitting 
\begin{equation}
\label{eq:vnsi}
\Delta H_m = V_{\rm NSI} =  V_d \sqrt{4\epsilon_{\mu\tau}^2 + \epsilon^{\prime2}}~. 
\end{equation}
The parameters $\sin2\xi$ and $V_{\rm NSI}$ determine the pattern of oscillations at very high energies, so that the transition probability equals 
\begin{equation}
\label{eq:prob0}
P(\nu_\mu \rightarrow \nu_\tau) =  \sin^2 2\xi~ \sin^2 \phi_{matt}~. 
\end{equation}
Here the matter half-phase, $\phi_{matt}$, equals  
\begin{equation}\label{eq:phimatt}
\phi_{matt} = \frac{\overline{V}_{\rm NSI} L}{2} = 
\frac{\overline{V}_d L}{2} 
\sqrt{4\epsilon_{\mu\tau}^2 + \epsilon^{\prime2}}~,   
\end{equation} 
where $\overline{V}_d$ is the averaged potential along the neutrino trajectory. (We assume that $\epsilon_{\mu\tau}$ and $\epsilon^{\prime}$ are constants.) Numerically, 
\begin{equation}
\phi_{matt} = 35 
\left(\frac{\bar{\rho}}{5.5\, {\rm g}~{\rm cm}^{-3}}\right)
\left(\frac{L}{2R_\oplus}\right)\sqrt{4\epsilon_{\mu\tau}^2 + \epsilon^{\prime2}}~,  
\end{equation}
where 
$$
\bar{\rho} = \frac{2}{L} \int_0^{L/2} dx \rho(x)~,
$$
is the average density along the neutrino trajectory and $R_\oplus$ is the Earth's radius. 

With decrease of $\epsilon_{\alpha \beta}$ the potential $V_{\rm NSI}$, and consequently, the phase $\phi_{matt}$ decrease. When $\phi_{matt} \ll 1$, we obtain from Eqs.~(\ref{eq:parameters}),~(\ref{eq:vnsi}) and (\ref{eq:phimatt}), the probability  
\begin{equation}
\label{eq:probapp}
P(\nu_\mu \rightarrow \nu_\tau) \approx  (\epsilon_{\mu\nu} \overline{V}_d L)^2~,
\end{equation}
which reproduces the result of~\cite{Akhmedov:2000cs}. It does not depend on $\epsilon^{\prime}$, so that high energy data restrict $\epsilon_{\mu\tau}$ and are insensitive to $\epsilon^{\prime}$. The same result is valid for both neutrino and antineutrino and for both signs of $\epsilon_{\mu\nu}$. 

The result in Eq.~(\ref{eq:probapp}) allows us to immediately estimate sensitivity of a given experiment to $\epsilon_{\mu\tau}$. From Eq.~(\ref{eq:probapp}) we have 
\begin{equation}
\epsilon_{\mu\nu} =  \frac{1}{\overline{V}_d L} \sqrt{P(\nu_\mu \rightarrow \nu_\tau)}~. 
\label{eq:sens}
\end{equation}
For atmospheric neutrinos the maximal value $\overline{V}_d L \approx 62$ corresponds to the trajectory along Earth's diameter, so that $\epsilon_{\mu\nu}^{\rm min} = 0.016\sqrt{P}$. For the $\sim 10\%$ accuracy of measurement of the probability, $P \approx 0.1$ (what we have now), we obtain $\epsilon_{\mu\nu}^{\rm min} = 5 \times 10^{-3}$. The accuracy $\sim 1\%$ (which can be considered as the ultimate one) leads to $\epsilon_{\mu\nu}^{\rm min} \sim 2 \times 10^{-3}$. The sensitivity in this region is restricted since the NSI effects are quadratic in strength $P\propto \epsilon_{\mu\nu}^2$.

Using parameters $\xi$ and $V_{\rm NSI}$, we can rewrite the Hamiltonian in Eq.~(\ref{eq:h2nu}) as 
\begin{equation}
\label{eq:h2nuf}
H_{2\nu} = \frac{\Delta m_{31}^2}{2E_\nu}  \left[U(\theta_{23}) \left(
\begin{array}{cc}
 0 & 0  \\
  0 & 1 
\end{array} \right) U^\dagger (\theta_{23}) + R_0 U(\xi) \left(
\begin{array}{cc}
0 & 0  \\
0 & 1  
\end{array} \right)U^\dagger(\xi) \right]~ .
\end{equation}
Here 
\begin{equation}
R_0 \equiv \frac{2 E_\nu V_{\rm NSI}}{\Delta m_{31}^2} = \sqrt{2} G_F n_d  
\sqrt{4\epsilon_{\mu\tau}^2 + \epsilon^{\prime2}} \frac{2E_\nu}{\Delta m^2_{31}}~,
\end{equation}
is the relative strength of matter and vacuum contributions, or is the splitting due to matter effect in units of the vacuum splitting. Numerically  
\begin{equation}\label{eq:r0number}
R_0 = 0.5  \left(\frac{\bar{\rho}}{5.5~ {\rm g}~{\rm cm}^{-3}} \right) 
\left(\frac{E_\nu}{{\rm GeV}}\right) \sqrt{4\epsilon_{\mu\tau}^2 + \epsilon^{\prime2}}~.
\end{equation}

Diagonalizing the Hamiltonian in Eq.~(\ref{eq:h2nuf}) we find the difference of instantaneous eigenvalues 
\begin{equation}
\Delta H_m =  \frac{\Delta m_{31}^2}{2E_\nu} R ~, 
\end{equation}
where the resonance factor $R$ equals  
\begin{equation}
R^2 = 1 + R_0^2 + 2R_0 \cos2(\theta_{23} - \xi) = [R_0 + \cos2(\theta_{23}- \xi)]^2 + \sin^2 2(\theta_{23}- \xi)~.   
\label{eq:rfactor}
\end{equation}
The mixing angle $\Theta_m$ is given by
\begin{equation}
\label{eq:modifiedangle}
\sin^2 2\Theta_m = \frac{1}{R^2}\left(\sin 2\theta_{23} + R_0 \sin 2\xi \right)^2~, 
\end{equation}
or explicitly, 
\begin{equation}
\label{eq:angleex}
\sin^2 2\Theta_m = \frac{\left(\sin 2\theta_{23} + R_0 \sin 2\xi \right)^2}{1 + R_0^2 + 2R_0 \cos2(\theta_{23}- \xi)}~.
\end{equation} 
The oscillation half-phase equals 
\begin{equation}
\Phi_m =  \Delta H_m \frac{L}{2} =  \left(\frac{\Delta m_{31}^2 L}{4E_\nu}\right) \left[1  + R_0^2 + 2 R_0 
\cos2(\theta_{23}- \xi) \right]^{1/2}. 
\label{eq:phase}
\end{equation}
It can be rewritten as 
\begin{equation}\label{eq:phase2}
\Phi_{m} = (\phi_{vac} + \phi_{matt} ) \sqrt{1 - \frac{2R_0}{(1 + R_0)^2} \left[1 - \cos2(\theta_{23}- \xi) \right] }~,
\end{equation}
where 
\begin{equation}
\phi_{vac} \equiv \frac{\Delta m_{31}^2 L}{4E_\nu} 
\end{equation}
is the vacuum oscillation half-phase. 

In the case of constant density, oscillation probabilities have standard expression with oscillation depth and length given by $\sin^2 2\Theta_m$ and $2\pi/\Delta H_m$:
\begin{equation}\label{eq:probformula} 
P(\nu_\mu\to\nu_\mu)=1 - \sin^2 2\Theta_m \sin^2 \left(\frac{\Delta m^2_{31}L}{4E_\nu}R\right)~.  
\end{equation}
If $\epsilon_{\mu\tau}$ changes the sign, $\epsilon_{\mu\tau} \rightarrow - \epsilon_{\mu\tau}$, in the above formulae we need to change $\xi \rightarrow - \xi$; for $\epsilon^{\prime} \rightarrow -\epsilon^{\prime}$, we replace $\cos2(\theta_{23}- \xi)  \rightarrow - \cos2(\theta_{23} + \xi)$. Changing the signs of both parameters $\epsilon_{\mu\tau}$ and $\epsilon^{\prime}$ is equivalent to $R_0 \rightarrow -R_0$. In the above formulae we assumed normal hierarchy of neutrino masses ($\Delta m^2_{31} > 0$). Inversion of the hierarchy can be described by $\Delta m^2_{31} \rightarrow - \Delta m^2_{31}$.

According to Eq.~(\ref{eq:rfactor}) the resonance condition, which ensures minimal value of the resonance factor, reads as
\begin{equation}
R_0 = - \cos2(\theta_{23}- \xi)~. 
\end{equation}
(For $\xi = 0$ it reduces to the usual MSW resonance condition.) Then the resonance energy equals to
\begin{equation}\label{eq:res}
E_R = - \frac{\Delta m_{31}^2}{2 V_{\rm NSI}} \cos2(\theta_{23}- \xi) = 
- \frac{\Delta m_{31}^2}{2 V_{d}\sqrt{4\epsilon_{\mu\tau}^2 + \epsilon^{\prime2}}} \cos2(\theta_{23}- \xi)~.
\end{equation}
For $\epsilon_{\alpha \beta} \sim 10^{-2}$, which is at the level of current sensitivity, we obtain $E_R \sim 100$~GeV. 

In resonance we have
$$
R^2 = \sin^2 2(\theta_{23}- \xi) = 1- R_0^2~,   
$$ 
and the mixing in Eq.~(\ref{eq:angleex}) becomes 
\begin{equation}\label{eq:resmix}
\sin^2 2\Theta_m = \cos^2 2\xi~. 
\end{equation}
For $\xi = 0$ this equation reproduces the usual result $\sin^2 2\Theta_m = 1$. In contrast, if the matter mixing is maximal, $\xi = \pi/4$, we obtain $\sin^2 2\Theta_m = 0$;  {\it i.e.}, mixing disappears in resonance. The oscillation phase in resonance equals 
\begin{equation}
\Phi_m  = \frac{\Delta m_{31}^2 L}{4E} \sin 2(\theta_{23} - \xi)~. 
\label{eq:phase3}
\end{equation}
The resonance is in the neutrino channels for negative $\epsilon$'s and in the antineutrino channels for positive $\epsilon$'s. 

Maximal interplay between the vacuum parameters and NSI is at energies $E \sim E_R$ or $R_0 \sim - 1$. The resonance energy increases with the decrease of $\epsilon$'s. In the range $E \simeq E_R$ the probability depends on $\epsilon$'s linearly due to interference with usual oscillations, as can be seen from expression for mixing angle in Eq.~(\ref{eq:modifiedangle}). 

The ratio $R_0$ quantifies the relative effect of NSI. For $R_0 \rightarrow 0$ (low energies) we have $\xi = 0$, $R \rightarrow 1$, and the parameters of oscillations reduce to the vacuum values:   
\begin{equation}
\sin^2 2\Theta_m = \sin 2\theta_{23}, ~~~~ 
\Delta H_m  \rightarrow \frac{\Delta m_{31}^2}{2E},~~~~
\Phi_m \rightarrow \phi_{vac}. 
\end{equation}

In the limit of high energies, that is, $R_0 \rightarrow \infty$ and $R \rightarrow R_0$, we reproduce the matter dominating results in Eqs.~(\ref{eq:parameters}) and (\ref{eq:vnsi}):   
\begin{equation}
\sin 2\Theta_m \rightarrow \sin 2\xi, ~~~  
\Delta H_m \rightarrow V_{\rm NSI}. 
\end{equation} 
With the decrease of $\epsilon$'s the energy where NSI effect becomes dominating increases, but the oscillation length increases as $l_m \propto 1/\Delta H_m \sim 1/\epsilon$. At very small $\epsilon$, the length $l_m$ becomes much larger than the diameter of the Earth and the oscillation phase becomes small $\phi_{matt} \ll 1$. In this case expanding the sine in the expression for the probability we obtain  
\begin{equation}
\label{eq:probformula} 
P(\nu_\mu\to\nu_\mu)=1 - \left(\sin 2\theta_{23} + R_0 \sin 2\xi \right)^2 
\cdot \left(\frac{\Delta m^2_{31}L}{4E_\nu}\right)^2~.  
\end{equation}
For $\xi = 0$ we would reproduce the ``vacuum mimicking'' result~\cite{Akhmedov:2000cs,Yasuda:2001va} for which the probability coincides with the vacuum oscillation probability in spite of the fact that the matter effect dominates ($R_0 > 1$). For $\xi \neq 0$ there is a deviation from the ``vacuum mimicking''. For $R_0 \gg 1$ we obtain from Eq.~(\ref{eq:probformula}) the expression in Eq.~(\ref{eq:probapp}).

Results for antineutrinos can be obtained by changing the sign of $R_0$:  $R_0 \rightarrow - R_0$, or equivalently by change of signs of all $\epsilon$'s (or $V_d$). 

\subsection{Probabilities for two extreme cases\label{sec:probc}}

Let us consider two extreme cases:

1).  Flavor off-diagonal NSI: $\epsilon_{\mu\tau} \neq0$, and $\epsilon^\prime \equiv \epsilon_{\tau\tau}-\epsilon_{\mu\mu}=0$, which corresponds to the universal NSI. In this case we have: $\sin 2\xi=1$. The mixing parameter equals 
\begin{equation}\label{eq:angleex1}
\sin^2 2\Theta_m = \frac{\left(\sin 2\theta_{23} + R_0 \right)^2}{1 + R_0^2 + 2R_0\sin 2\theta_{23}} = 
\frac{1}{1 + \cos^2 2\theta_{23} (R_0 + \sin 2\theta_{23})^{-2}}~, 
\end{equation}
and it converges to maximal one for large $R_0$. The resonance factor becomes 
\begin{equation}
R^2 = [R_0 + \sin 2 \theta_{23}]^2 + \cos^2 2\theta_{23}~. 
\end{equation}

In resonance the phase is suppressed by the factor $\cos 2\theta_{23}$:
\begin{equation}
\Phi_m  =\frac{\Delta m_{31}^2 L}{4E_\nu} \cos 2 \theta_{23}~,
\label{eq:phase4}
\end{equation}
(see Eq.~(\ref{eq:phase3})) and it is zero for the maximal 2-3 mixing. Far from the resonance ($R_0 \neq - 1$), according to Eq.~(\ref{eq:phase2}) the total oscillation phase equals approximately to the sum of matter and vacuum phases:
\begin{equation}
\label{eq:phase1}
\Phi_m \approx  \phi_{vac} + \phi_{matt} =  \frac{\Delta m^2_{32}L(1+R_0)}{4E_\nu} = 
\frac{\Delta m^2_{32}L}{4E_\nu}+ V_d L\epsilon_{\mu\tau}~,
\end{equation}
so that modification of the phase is independent of neutrino energy. In the high energy regime, $E_\nu\gtrsim100$~GeV, the standard contribution to the phase (the first term in Eq.~(\ref{eq:phase1})) becomes negligible, whereas the second one equals
\begin{equation}
\phi_{matt} = 70 \left(\frac{\bar{\rho}}{5.5~ {\rm g}~{\rm cm}^{-3}}\right) 
\left(\frac{L}{2R_\oplus}\right) \epsilon_{\mu \tau}~. 
\end{equation}
For neutrinos crossing the center of Earth, $\cos\theta_z=-1$, this equation gives $\phi_{matt} = 62\epsilon_{\mu\tau}$. The phase equals $\pi/2$, so that the muon survival probability reaches its minimum when $\epsilon_{\mu\tau} = 2.5\times10^{-2}$. For $\epsilon_{\mu\tau}\gtrsim2.5\times10^{-2}$ the minimum of probability occurs at $\cos\theta_z>-1$. For example, for $\epsilon_{\mu\tau} \simeq 0.05$, the probability $P(\nu_\mu\to\nu_\mu)\simeq0$ at $\cos\theta_z\simeq-0.8$. In asymptotics for small $\epsilon_{\mu\tau}$ we have according to Eq.~(\ref{eq:probapp}) $P = \phi_{matt}^2$.

\begin{figure}[h!]
\centering
\subfloat[]{
\includegraphics[width=0.5\textwidth]{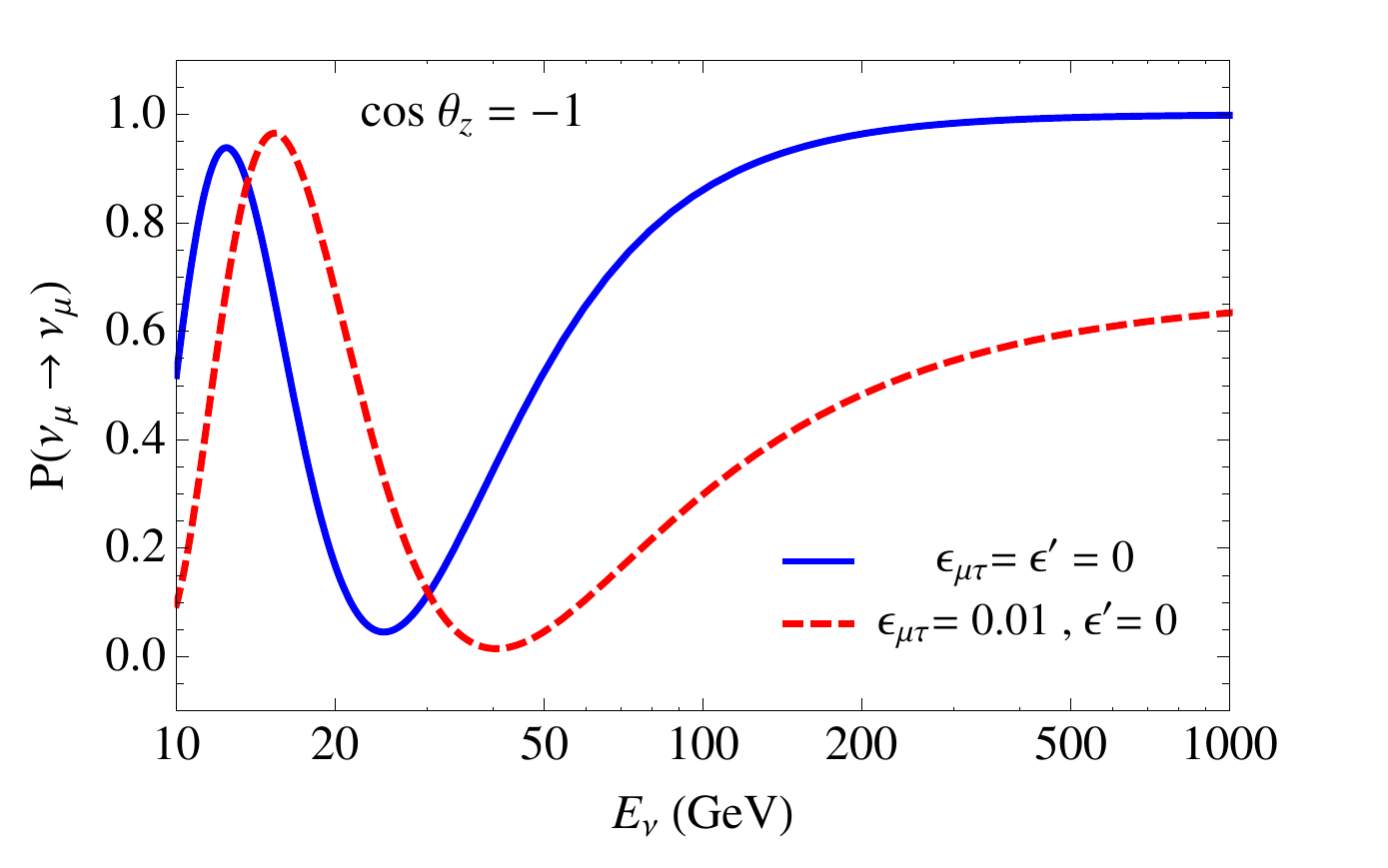}
 \label{fig:muwholea}
}
\subfloat[]{
\includegraphics[width=0.5\textwidth]{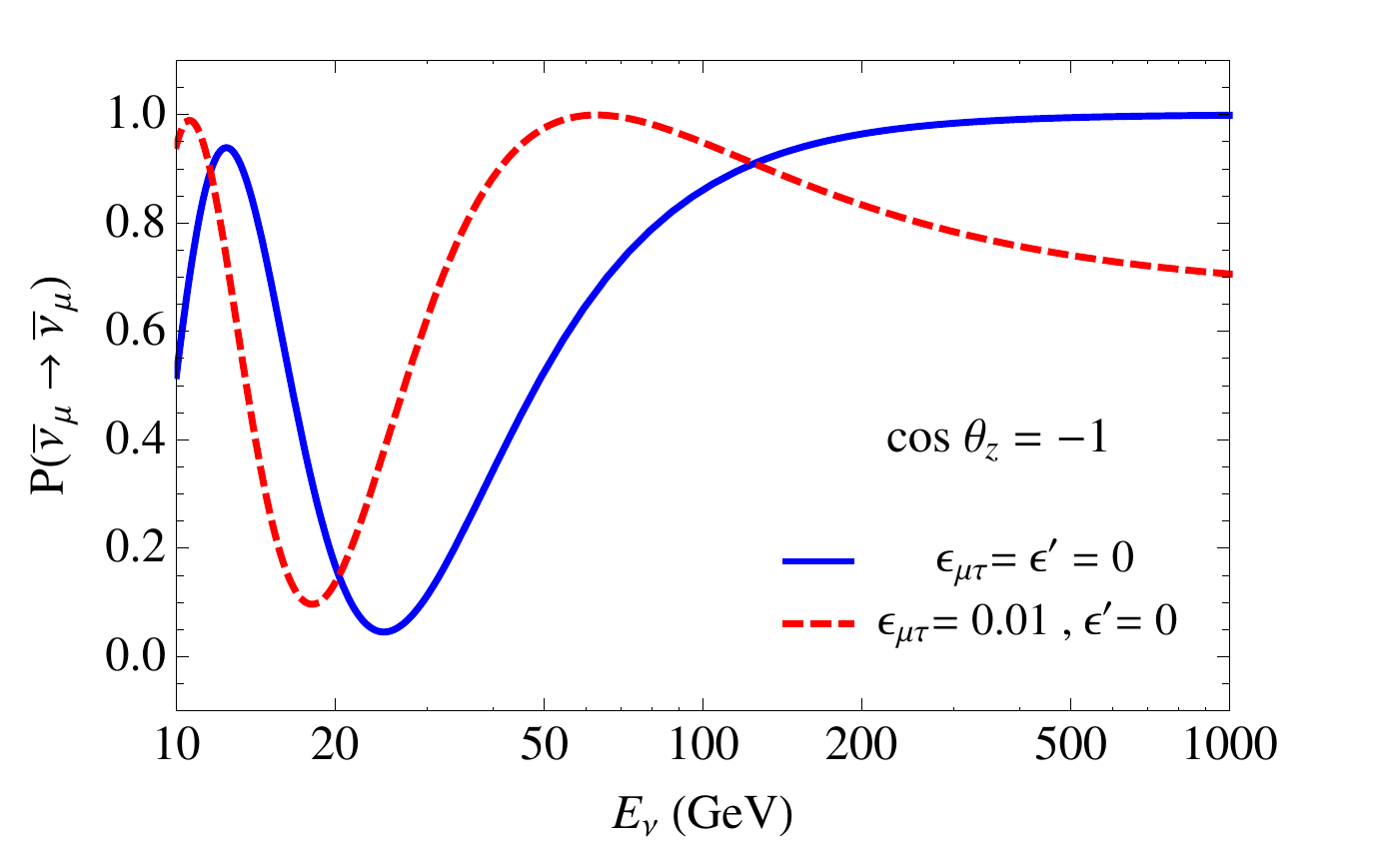}
 \label{fig:mubarwhole}
}
\caption{\label{fig:NSIwhole}The oscillation probabilities  $P({\nu}_\mu\to{\nu}_\mu)$ (a) and 
$P(\bar{\nu}_\mu\to\bar{\nu}_\mu)$ (b) as functions of neutrino  energy for $\cos\theta_z=-1$. The dashed red curve is for NSI with parameters $\epsilon_{\mu\tau}=0.01$ and $\epsilon^\prime=0$,  and the solid blue curve is for the standard $3\nu$ oscillations.}
\end{figure}

In Fig.~\ref{fig:NSIwhole} we show the $\nu_\mu-$ and $\bar{\nu}_\mu-$ survival probabilities as functions of the neutrino energy with and without NSI. The plots correspond to opposite signs of $\epsilon_{\mu\tau}$. According to this figure and our analytical consideration at low energies, $E_\nu \lesssim 100$~GeV, NSI lead to shift of the oscillatory pattern to larger ($\nu$) or smaller ($\bar{\nu}$) energies for $\epsilon_{\mu\tau} > 0$. Indeed, for $\cos\theta_z=-1$ the first minimum of the $\nu_\mu\to\nu_\mu$ vacuum oscillation probability ($\phi_{vac} = \pi/2$) occurs at $E_\nu \sim25$~GeV. The second term of Eq.~(\ref{eq:phase1}) leads to a shift of the minimum to higher or lower energies, depending on the sign of $\epsilon_{\mu\tau}$. The mixing angle also modifies in the presence of NSI and the depth of minimum changes, although the change is small.

In the case of $\epsilon_{\mu\tau} > 0$ and normal mass hierarchy the resonance is in the $\bar{\nu}$ channel ($V_d < 0$) and the resonance energy equals, according to Eq.~(\ref{eq:res}),
\begin{equation}
E_R = - \frac{\Delta m_{31}^2}{4 \epsilon_{\mu\tau}\overline{V}_d(\theta_z)} \sin 2 \theta_{23}~.  
\label{eq:res2}
\end{equation}
For $\cos\theta_z=-1$ we obtain $E_R \approx 60$~GeV. In resonance, $R_0 = - \sin\theta_{23}$, the mixing is zero, $\sin^2 2\Theta_m = 0$, and oscillation effects are absent. This corresponds to $P(\bar{\nu}_\mu\to\bar{\nu}_\mu) = 1$ at $E \approx 60$~GeV in Fig.~\ref{fig:mubarwhole}. With the increase of energy both neutrino and antineutrino probabilities in Fig.~\ref{fig:NSIwhole} converge to the same asymptotic value $\sim \phi_{matt}^2$.

\begin{figure}[t!]
\centering
\subfloat[]{
 \includegraphics[trim= 0mm 0mm 0mm 
80mm,clip,width=0.5\textwidth]{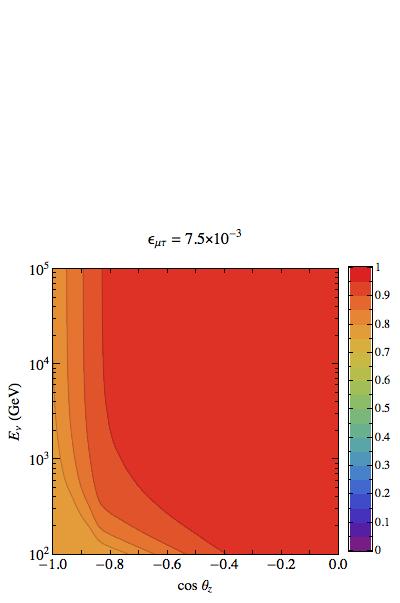}
  \label{fig:prob,29,21,high}
}
\subfloat[]{
 \includegraphics[trim= 0mm 0mm 0mm 
80mm,clip,width=0.5\textwidth]{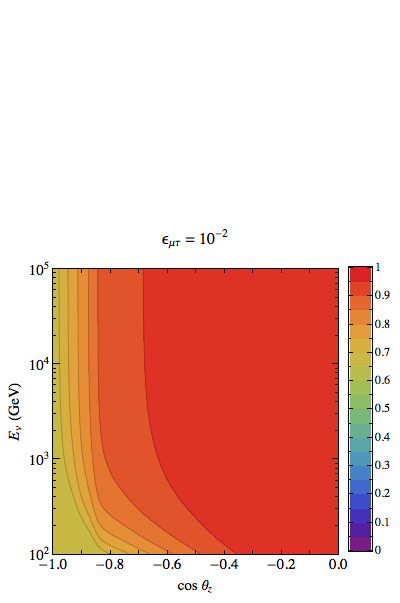}
  \label{fig:prob,34,21,high}
}
\caption{\label{fig:probhigh}The survival probability $P(\nu_\mu\to\nu_\mu)$ in the presence of NSI in the high energy range $E_\nu > 100$~GeV for two different values of $\epsilon_{\mu\tau}$. We take $\epsilon^\prime=0$.}
\end{figure}

\begin{figure}[t!]
\centering
\subfloat[]{
 \includegraphics[trim= 0mm 0mm 0mm 
80mm,clip,width=0.5\textwidth]{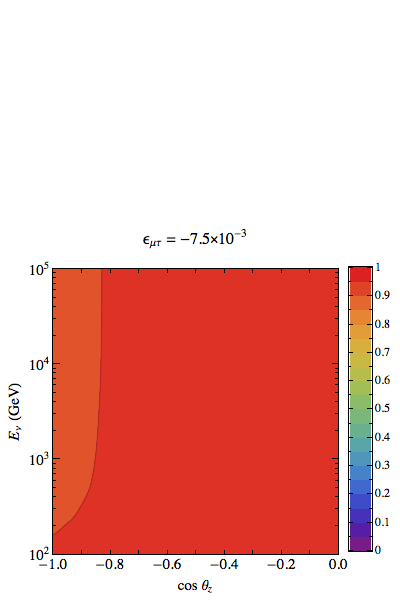}
  \label{fig:prob,13,21,high}
}
\subfloat[]{
 \includegraphics[trim= 0mm 0mm 0mm 
80mm,clip,width=0.5\textwidth]{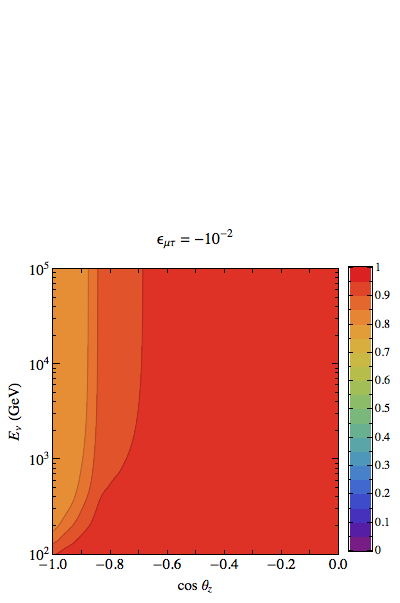}
  \label{fig:prob,8,21,high}
}
\caption{\label{fig:probhighnegative}
The same as  in Fig.~\ref{fig:probhigh} but for negative values of $\epsilon_{\mu\tau}$ (or for antineutrinos).}
\end{figure}

Figs.~\ref{fig:probhigh} and \ref{fig:probhighnegative} show oscillograms for the $\nu_\mu$ survival probability in the high energy range ($E_\nu\gtrsim 100$~GeV) for $\epsilon^\prime=0$ and different values of $\epsilon_{\mu\tau}$. The resonance energy is in the interval $(80 - 200)$~GeV. The resonance is realized for negative values of $\epsilon_{\mu\tau}$ (Fig.~\ref{fig:probhighnegative}). According to Eq.~(\ref{eq:resmix}) in resonance (due to $\xi = \pi/4$) the mixing becomes zero $\sin^2 2\Theta_m = 0$. Furthermore, the matter and vacuum phases have opposite signs and cancel each other. As a result, the oscillations are strongly suppressed at $E_\nu \sim E_R$ in agreement with our analytical consideration. Above resonance the matter dominated oscillations are realized and at $E_\nu \gg E_R$ both signs of $\epsilon_{\mu\tau}$ give the same results. 
  
Notice that at high energies usual vacuum oscillations are strongly suppressed, so that from these figures one infers immediately the difference of probabilities with and without NSI. 

\begin{figure}[t!]
\centering
\subfloat[]{
 \includegraphics[trim= 0mm 0mm 0mm 
80mm,clip,width=0.5\textwidth]{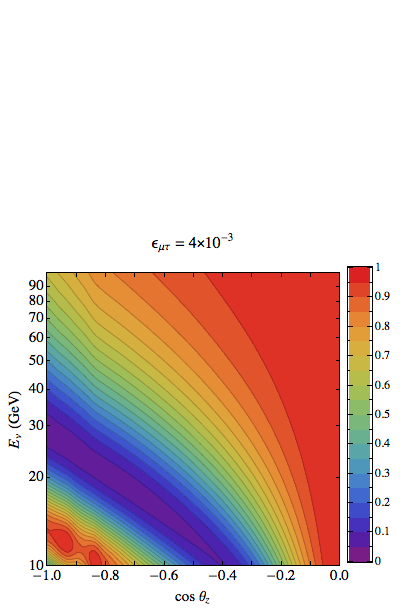}
  \label{fig:prob,29,21,low}
}
\subfloat[]{
 \includegraphics[trim= 0mm 0mm 0mm 
80mm,clip,width=0.5\textwidth]{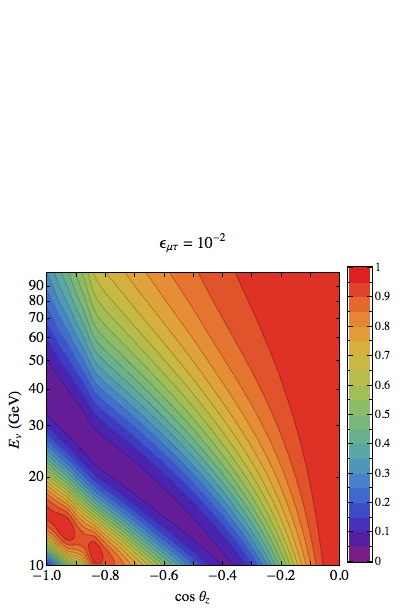}
  \label{fig:prob,34,21,low}
}
\quad
\subfloat[]{
 \includegraphics[trim= 0mm 0mm 0mm 
165mm,clip,width=0.5\textwidth]{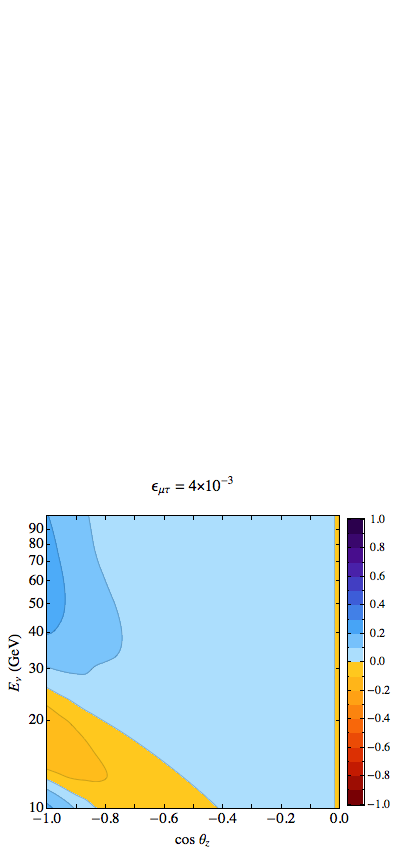}
  \label{fig:probdiff,29,21,low}
}
\subfloat[]{
 \includegraphics[trim= 0mm 0mm 0mm 
165mm,clip,width=0.5\textwidth]{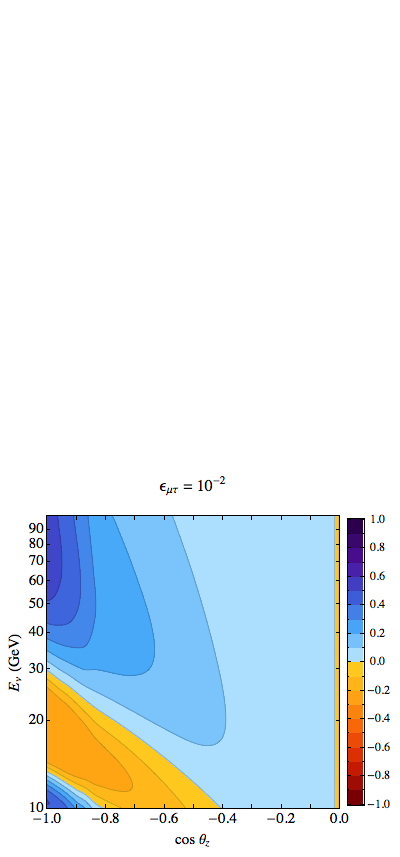}
  \label{fig:probdiff,34,21,low}
}
\caption{
\label{fig:problow1} 
The oscillograms for the survival probability $P(\nu_\mu\to\nu_\mu)$ in the presence of NSI (panels a and b) and the difference of probabilities with and without NSI (panels c, d) [see Eq.~(\ref{eq:probdiff})] in the low energy range $10~{\rm GeV}<E_\nu<100$~GeV for two different values of $\epsilon_{\mu\tau}$. We take $\epsilon^\prime=0$.}
\end{figure}

\begin{figure}[t!]
\centering
\subfloat[]{
 \includegraphics[trim= 0mm 0mm 0mm 
80mm,clip,width=0.5\textwidth]{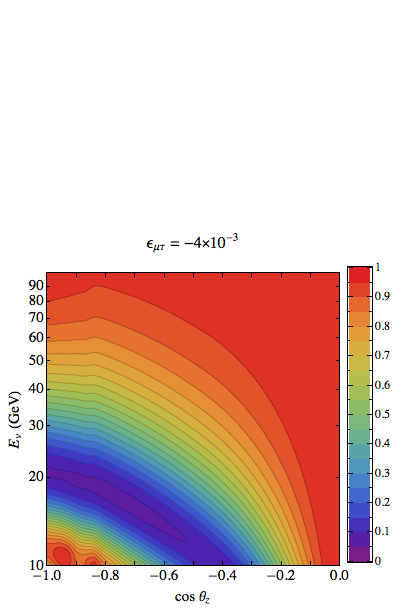}
  \label{fig:prob,13,21,low}
}
\subfloat[]{
 \includegraphics[trim= 0mm 0mm 0mm 
80mm,clip,width=0.5\textwidth]{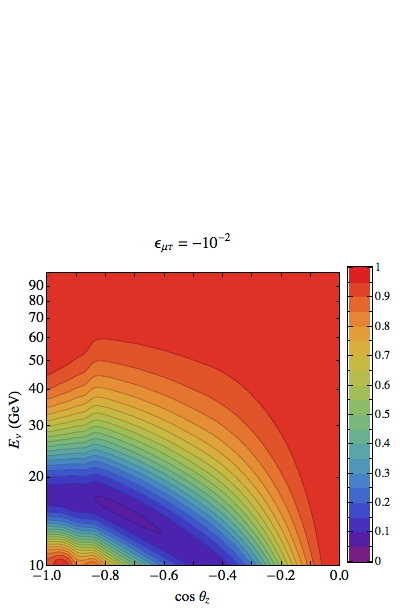}
  \label{fig:prob,8,21,low}
}
\quad
\subfloat[]{
 \includegraphics[trim= 0mm 0mm 0mm 
165mm,clip,width=0.5\textwidth]{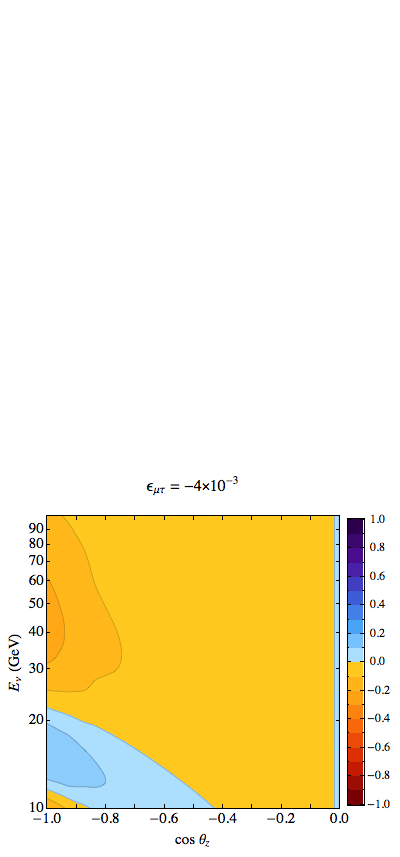}
  \label{fig:probdiff,13,21,low}
}
\subfloat[]{
 \includegraphics[trim= 0mm 0mm 0mm 
165mm,clip,width=0.5\textwidth]{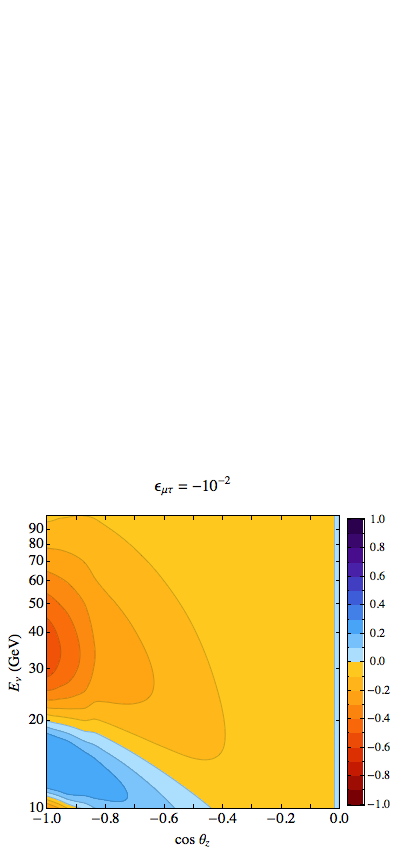}
  \label{fig:probdiff,8,21,low}
}
\caption{\label{fig:problow2}The same as in Fig.~\ref{fig:problow1} for negative values of $\epsilon_{\mu\tau}$ (or for antineutrinos).}
\end{figure}

In Figs.~\ref{fig:problow1} and \ref{fig:problow2} we show the $\nu_\mu$ survival probability patterns in the low energy range $10~{\rm GeV}<E_\nu<100$~GeV for different values of $\epsilon_{\mu\tau}$. In this energy range, as expected from the analytical consideration in Eqs.~(\ref{eq:probformula}) - (\ref{eq:phase1}), the NSI's lead to shift of the oscillatory pattern and in particular, minimum of $\nu_\mu$ survival probability. The positive (negative) values of $\epsilon_{\mu\tau}$ shift the pattern of $P(\nu_\mu\to\nu_\mu)$ to higher (lower) energies: $E_\nu^{min} \simeq 24$~GeV, $30$~GeV and $38$~GeV for $\epsilon_{\mu\tau} = 0,~ 4 \times 10^{-3}$ and $10^{-2}$ correspondingly. The depth of the minimum in $P(\nu_\mu\to\nu_\mu)$ decreases for negative values of $\epsilon_{\mu\tau}$. To illustrate the effect of NSI on the oscillation pattern, we also show in Figs.~\ref{fig:problow1} and \ref{fig:problow2} the difference of probabilities
\begin{equation}
\label{eq:probdiff}
P^{\rm STD}(\nu_\mu \to \nu_\mu) - P(\nu_\mu \to \nu_\mu) (\epsilon_{\mu\tau},\epsilon^\prime=0)~,  
\end{equation}
where $P^{\rm STD}$ is the oscillation probability in the $3\nu$ standard framework without NSI. 
Maximal difference is in the range $40 - 100$~GeV, as well as at low energies $15 - 20$~GeV. 

As follows from the Figs.~\ref{fig:problow1} and \ref{fig:problow2} (bottom panels) interference of the NSI and usual oscillation effects has opposite signs in different kinematical $(E_\nu - \cos \theta_z)$ regions. Therefore to enhance the sensitivity to the NSI parameters one should avoid integration over regions of opposite signs. Comparing bottom panels in Figs.~\ref{fig:problow1} and \ref{fig:problow2} we see also that the NSI effects have opposite signs for neutrinos and antineutrinos, and therefore summation of the $\nu$ and $\bar{\nu}$ signals leads to partial (due to difference of the cross-sections and fluxes) cancellation of effects. Separation of the neutrino and antineutrino signals would allow to improve the sensitivity.\\

2).  Flavor conserving NSI: $\epsilon_{\mu\tau}=0$, $\epsilon^\prime\equiv \epsilon_{\tau\tau}-\epsilon_{\mu\mu}\neq0$. Now NSI are flavor non-universal. In this case $\sin 2\xi=0$, and for the mixing and mass splitting we obtain the usual MSW formulas with the potential given by $V_{\rm NSI} = V_d \epsilon^{\prime}$:   
\begin{equation}
\label{eq:angleex2}
\sin^2 2\Theta_m = 
\frac{\sin^2 2\theta_{23}}{
(R_0 + \cos2 \theta_{23})^2 + \sin^2 2\theta_{23}}~, 
\end{equation}
and 
\begin{equation}
\Delta H_m  = 
\frac{\Delta m_{31}^2}{2E}  \left[ 
(R_0 + \cos2 \theta_{23})^2 + \sin^2 2\theta_{23} 
\right]^{1/2}. 
\end{equation}
Consequently, the oscillation probabilities will have the standard MSW dependences on the neutrino energy. The parameter $R_0$ (from Eq.~(\ref{eq:r0number})) takes the value 
\begin{equation}
R_0 = 0.5 \left(\frac{\bar{\rho} (\theta_z) }{5.5~ {\rm g}~{\rm cm}^{-3}} \right)
\left(\frac{E_\nu}{{\rm GeV}}\right)\epsilon^\prime~.
\end{equation}
The resonance condition reads $R_0  = - \cos2 \theta_{23}$, and the resonance energy is 
\begin{equation}
E_R = 2~ {\rm GeV} 
\left(\frac{5.5 ~{\rm g}~{\rm cm}^{-3}}{\bar{\rho} (\theta_z) } \right)
\left(\frac{\cos2 \theta_{23}}{\epsilon^\prime}
\right) .
\end{equation}
For $\epsilon^\prime = 5 \times 10^{-2}$ we obtain $E_R \sim 10$ GeV for the mantle crossing trajectories. Resonance enhancement of oscillation is very weak since the vacuum mixing is already large. Therefore the main effect of NSI is the suppression of oscillations at energies above the resonance $E_\nu > E_R$. In non-resonance channel the suppression starts even at lower energies. With decrease of $\epsilon^\prime$ the region of small survival probability shifts to higher energies as $E_\nu \propto 1/\epsilon^\prime$.

The oscillation phase, Eq.~(\ref{eq:phase}), for neutrinos crossing the center of Earth can be approximated as
\begin{equation}
\Phi_m =  38 \left( \frac{{\rm GeV}}{E_\nu}\right) \sqrt{1 +     
\cos 2 \theta_{23}\left( \frac{E_\nu}{{\rm GeV}}\right) \epsilon^{\prime} + 
0.25\left( \frac{E_\nu}{{\rm GeV}}\right)^2 \epsilon^{\prime 2}}~.
\end{equation}
The mixing angle equals
\begin{equation}
\label{eq:primeangle}
\sin^2 2\Theta_m = \frac{ \sin^2 2\theta_{23}}{1+    
\cos 2 \theta_{23}\left( \frac{E_\nu}{{\rm GeV}}\right) \epsilon^{\prime} + 
0.25\left( \frac{E_\nu}{{\rm GeV}}\right)^2 \epsilon^{\prime 2}}~. 
\end{equation}
Since $\xi = 0$, at high energies ($E_\nu \gtrsim E_R$) the oscillation phase is small and the vacuum mimicking is realized with usual vacuum oscillation result.   

Let us estimate sensitivity to $\epsilon^\prime$. Here, in contrast to very high energies, the effect is linear in $\epsilon^\prime$ due to interference with usual oscillations, although the linear terms appear in combination with the small factor $\cos 2\theta_{23}$: 
$$
\delta \equiv 2 \cos 2\theta_{23} R_0 =  
\frac{2\epsilon^{\prime} \cos 2\theta_{23} 2E_\nu V_d}{\Delta m^2_{31}}~. 
$$ 
Thus, the linear terms appear only if there is a deviation of the 2-3 mixing from maximal. We assume that $\epsilon^{\prime 2}$ terms are negligible which is realized for small enough $\epsilon^{\prime}$. Then in the lowest approximation in $\delta$, the change of oscillation probability due to NSI equals 
\begin{equation}  
\Delta P \approx \delta \left(- \sin^2 \phi_{vac} + 
\phi_{vac} \sin \phi_{vac} \cos \phi_{vac} \right)~.
\end{equation}
The sensitivity can be estimated as
\begin{equation}
\epsilon^{\prime}  \sim \Delta P  \frac{\Delta m_{31}^2}{2 E_\nu V_d} \left(- \sin^2 \phi_{vac} +
\phi_{vac} \sin \phi_{vac} \cos \phi_{vac} \right)^{-1}. 
\end{equation}
The factor in brackets, which depends on the oscillation phase in vacuum, is typically smaller than 1. Therefore for $\Delta P = 0.1$ and $E_\nu = 30$~GeV we obtain $\epsilon^{\prime} \sim 2 \times 10^{-2}$. 

At high energies, $E_\nu > E_R$, the mixing is strongly suppressed, $\sin^2 2\Theta_m \approx R_0^{-2} \sin^2 2\theta_{23}$, and consequently, the oscillation effects vanish.

\begin{figure}[t!]
\centering
\subfloat[]{
 \includegraphics[trim= 0mm 0mm 0mm 
80mm,clip,width=0.5\textwidth]{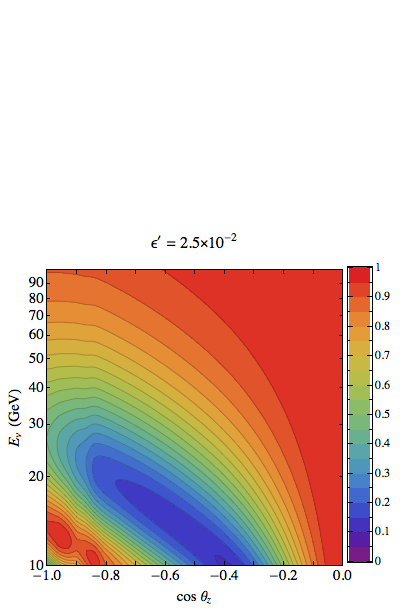}
  \label{fig:prob,29,21,low}
}
\subfloat[]{
 \includegraphics[trim= 0mm 0mm 0mm 
80mm,clip,width=0.5\textwidth]{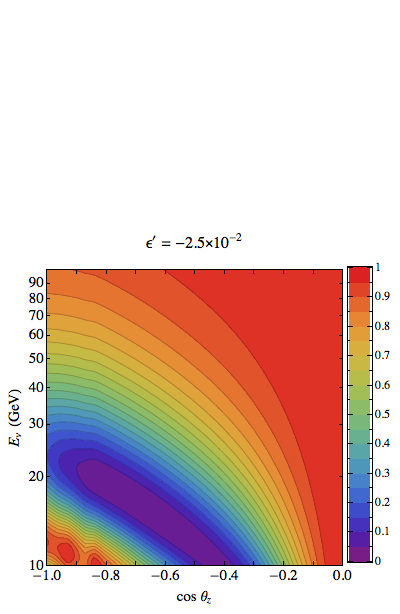}
  \label{fig:prob,34,21,low}
}
\quad
\subfloat[]{
 \includegraphics[trim= 0mm 0mm 0mm 
165mm,clip,width=0.5\textwidth]{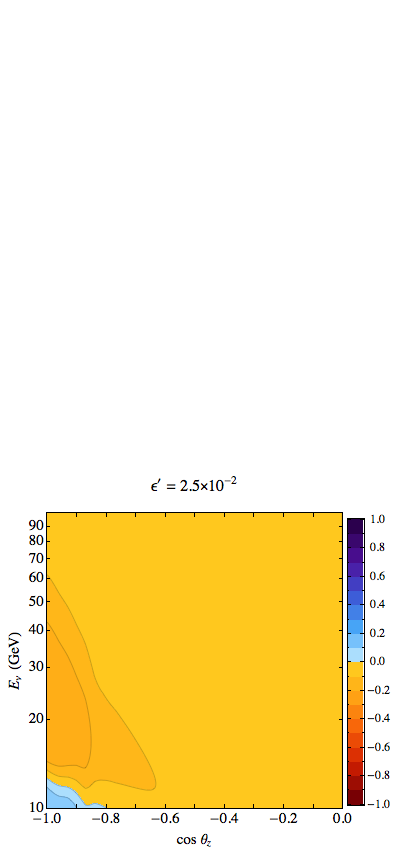}
  \label{fig:probdiff,29,21,low}
}
\subfloat[]{
 \includegraphics[trim= 0mm 0mm 0mm 
165mm,clip,width=0.5\textwidth]{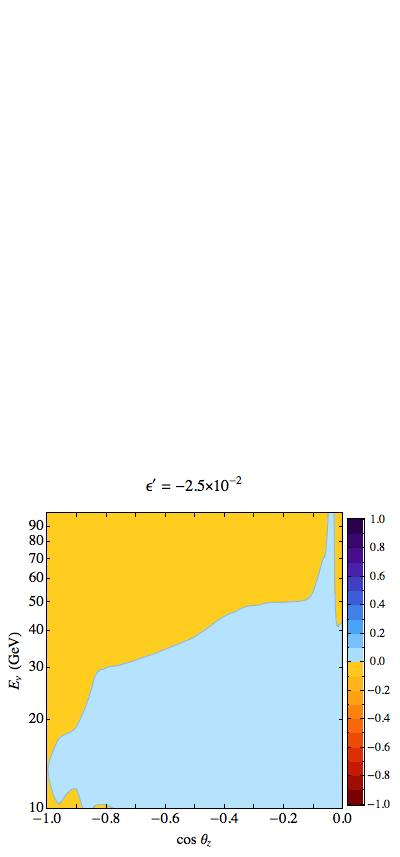}
  \label{fig:probdiff,34,21,low}
}
\caption{\label{fig:problowprime} The same as in Fig.~\ref{fig:problow2} for two  different values of 
$\epsilon^{\prime}$ and $\epsilon_{\mu\tau} = 0$. 
}
\end{figure}

Fig.~\ref{fig:problowprime} shows the oscillograms for $P(\nu_\mu\to\nu_\mu)$ and difference of oscillograms with and without NSI in the low energy range for $\epsilon^\prime = \pm 2.5\times 10^{-2}$ and $\epsilon_{\mu\tau}=0$. Recall that change of the sign of $\epsilon^\prime$ is equivalent to transition from neutrinos to antineutrinos. According to Fig.~\ref{fig:problowprime} the range of the strong $\epsilon^\prime$ effect is at $(20 - 40)$~GeV and it is stronger for positive values of $\epsilon^\prime$.

\subsection{Other contribution to the $\nu_\mu$ events}

The $\nu_e \rightarrow \nu_\mu$ contribution to the $\nu_\mu$ events is strongly suppressed at energies $E_\nu \gtrsim~20$~GeV, since: (i) the original flux of electron neutrinos becomes substantially -- a factor 5 -- smaller than the muon neutrino flux; (ii) the transition probability $P(\nu_e \rightarrow \nu_\mu)$ is strongly suppressed by usual matter effect being smaller than $1\%$. 

Another contribution to the $\nu_\mu$ events is from the $\nu_{\tau}$ flux formed by the $\nu_\mu \rightarrow \nu_\tau$ oscillations. The tau neutrinos produce tau leptons: $\nu_\tau + N \rightarrow \tau + X$ and in $17 \%$ cases $\tau$'s decay as $\tau \rightarrow \nu_\tau + \nu_\mu + \mu$. In average, muon takes 1/3 of the original energy of $\tau$, and therefore to reproduce the same configuration of the $\nu_\mu$ event the original $\nu_\tau$ energy should be 3 times larger than the $\nu_\mu$ energy. The corresponding flux is then 10 times smaller. Thus, the $\nu_{\tau}$ 
events give about $1- 2 \%$ contribution.

\section{Constraining $\epsilon_{\mu\tau}$ and $\epsilon^\prime$}
\label{sec:cons}

Let us  find constrains on the NSI parameters $\{\epsilon\}\equiv\{\epsilon_{\mu\tau},\epsilon^\prime \}$ from the atmospheric neutrino data collected by the IceCube-79 and DeepCore experiments. In subsection~\ref{sec:deepcore} we estimate the sensitivity of future DeepCore results.

\subsection{Zenith angle distributions of events}

In our analysis we use the atmospheric neutrino data collected by IceCube detector and named as the ``low" and ``high" energy samples~\cite{Gross:2013iq,sullivan}. The high energy sample is recorded by the IceCube detector with 79 strings and corresponds to the energy interval $100~{\rm GeV}-10~{\rm TeV}$. The ``low'' energy sample, $(20 -100)$~GeV, is composed of events detected by the DeepCore part of IceCube detector~\cite{Collaboration:2011ym}. 

Using the probabilities $P(\nu_\mu \rightarrow \nu_\mu)(\{\epsilon\})$ and $P(\nu_e \rightarrow  
\nu_\mu)(\{\epsilon\})$ discussed in the previous section we compute the expected zenith angle distribution of events in the presence of NSI. The number of events in the $i$-th zenith angle bin equals 
\begin{eqnarray}\label{eq:numbev}
& & \!\!\!\!\!\!\!\!\!\!\!\!\!\!\!\!\!\!\! N_i^{\rm exp}(\{\epsilon\})_{\rm low, ~high} = T\Delta\Omega \int_{\Delta_i \cos\theta_z} d\cos\theta_z \int_{\rm low, ~high} dE_\nu   \\
& &  \!\!\!\!\!\!\! \left[ \Phi_{\nu_\mu}(E_\nu,\cos\theta_z) P(\nu_\mu \rightarrow \nu_\mu) (\{\epsilon\}) 
+ \Phi_{\nu_e}  (E_\nu,\cos\theta_z) P(\nu_e \rightarrow  \nu_\mu) (\{\epsilon\}) \right] 
A_{\rm eff}^{\nu_\mu} (E_\nu,\cos\theta_z) \nonumber \\ \nonumber
& & + (\nu_\mu \to \bar{\nu}_\mu )~, \nonumber
\end{eqnarray} 
where $\Phi_{\nu_\mu(\bar{\nu}_\mu)}$ and $\Phi_{\nu_e(\bar{\nu}_e)}$ are the $\nu_\mu ~(\bar{\nu}_\mu)$ and $\nu_e~(\bar{\nu}_e)$ fluxes of atmospheric neutrinos~\cite{Honda:2006qj,Athar:2012it} and $A_{\rm eff}(E_\nu,\cos\theta_z)$ is the effective area. In Eq.~(\ref{eq:numbev}) $\Delta \Omega=2\pi$ is the azimuthal angular acceptance of the IceCube detector and $T$ is the live-time of detector. We take 10 bins in $\cos\theta_z$ with width 0.1. The number of events in the $i$-th bin is given by integration over the size of bin $\Delta_i \cos\theta_z$.

In the high energy range effects of the standard neutrino oscillation are almost absent, whereas the low energy sample covers the first minimum of $\nu_\mu$ survival probability. The experimental data are in a very good agreement with these expectations which allowed to claim observation of the atmospheric neutrino oscillation with $\sim5.6~\sigma$ significance~\cite{Gross:2013iq}. No deviation from the standard oscillation picture has been found and we analyze the ``low" and ``high" data samples to constrain the NSI parameters.

Since the effective area $A_{\rm eff}^{\nu_\mu} (E_\nu,\cos\theta_z)$ is not published yet, we estimated it by re-weighting the known effective area of IceCube-40~\cite{Esmaili:2012nz} such that the simulated zenith distribution of the IceCube-79 events is reproduced. As a cross-check, we also reproduced approximately the exclusion plot for the standard oscillation parameters $(\Delta m_{31}^2,\sin^22\theta_{23})$ as in~\cite{Gross:2013iq}. Although our limits, due to the cross-checks, are expected to be close to realistic ones, they should be considered as provisional and small changes are possible. 

\begin{figure}[t!]
\centering
\subfloat[]{
 \includegraphics[width=0.5\textwidth]{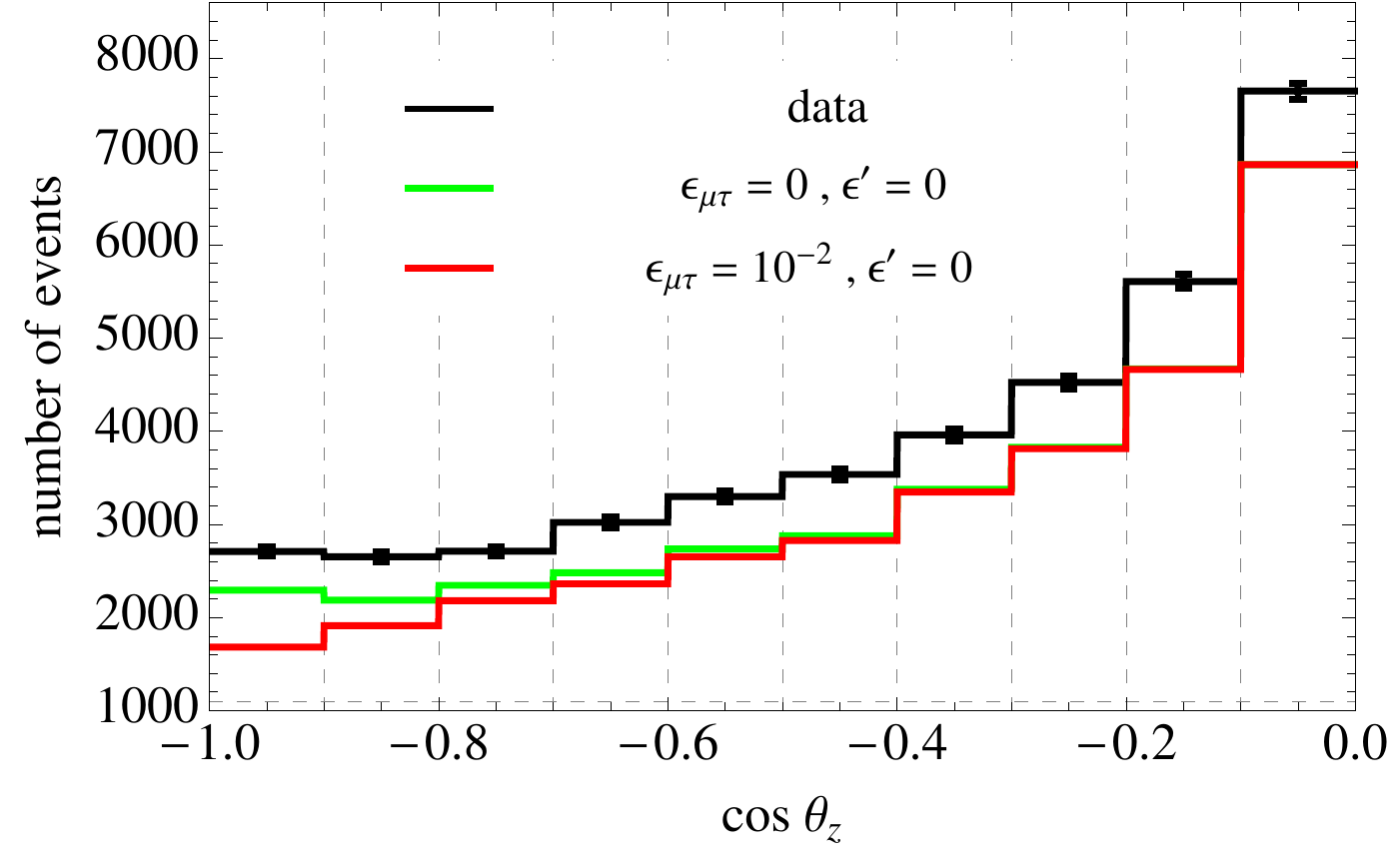}
 \label{fig:high6}
}
\subfloat[]{
 \includegraphics[width=0.5\textwidth]{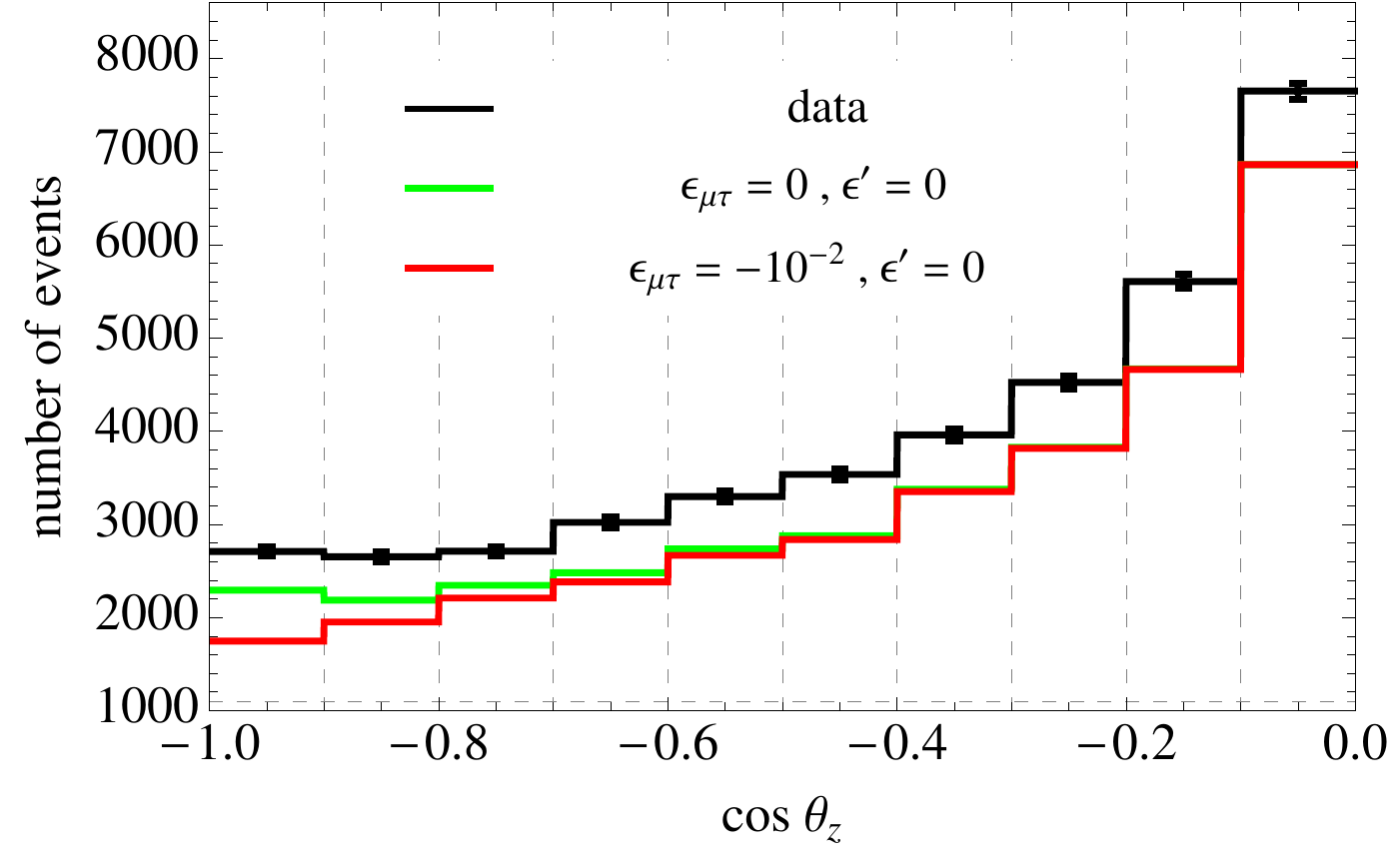}
 \label{fig:high16}
}
\caption{\label{fig:high616}The predicted zenith angle distribution of the high energy $\nu_\mu-$events at IceCube with (red) and without (green) NSI. For NSI parameters we use (a) $\epsilon_{\mu\tau}=0.01$, (b) $\epsilon_{\mu\tau}= - 0.01$, and $\epsilon^\prime=0$. The IceCube-79 data is shown by the black histogram.}
\end{figure}

In Fig.~\ref{fig:high616} we show the zenith angle distribution of events with and without NSI for the IceCube detector (high energy data). The distribution with NSI has been computed for $\epsilon_{\mu\tau} =\pm 0.01$. We show also the data points from IceCube-79~\cite{Gross:2013iq}. The distribution without NSI is (up to an overall normalization) in very good agreement with data. The NSI lead to additional suppression of number of events in vertical (upward going) directions, $\cos \theta_z < - 0.7$, and the corresponding distribution fits the data worse than the one without NSI. The distributions do not depend practically on the sign of $\epsilon_{\mu\tau}$. 

\begin{figure}[t!]
\centering
\subfloat[]{
 \includegraphics[width=0.5\textwidth]{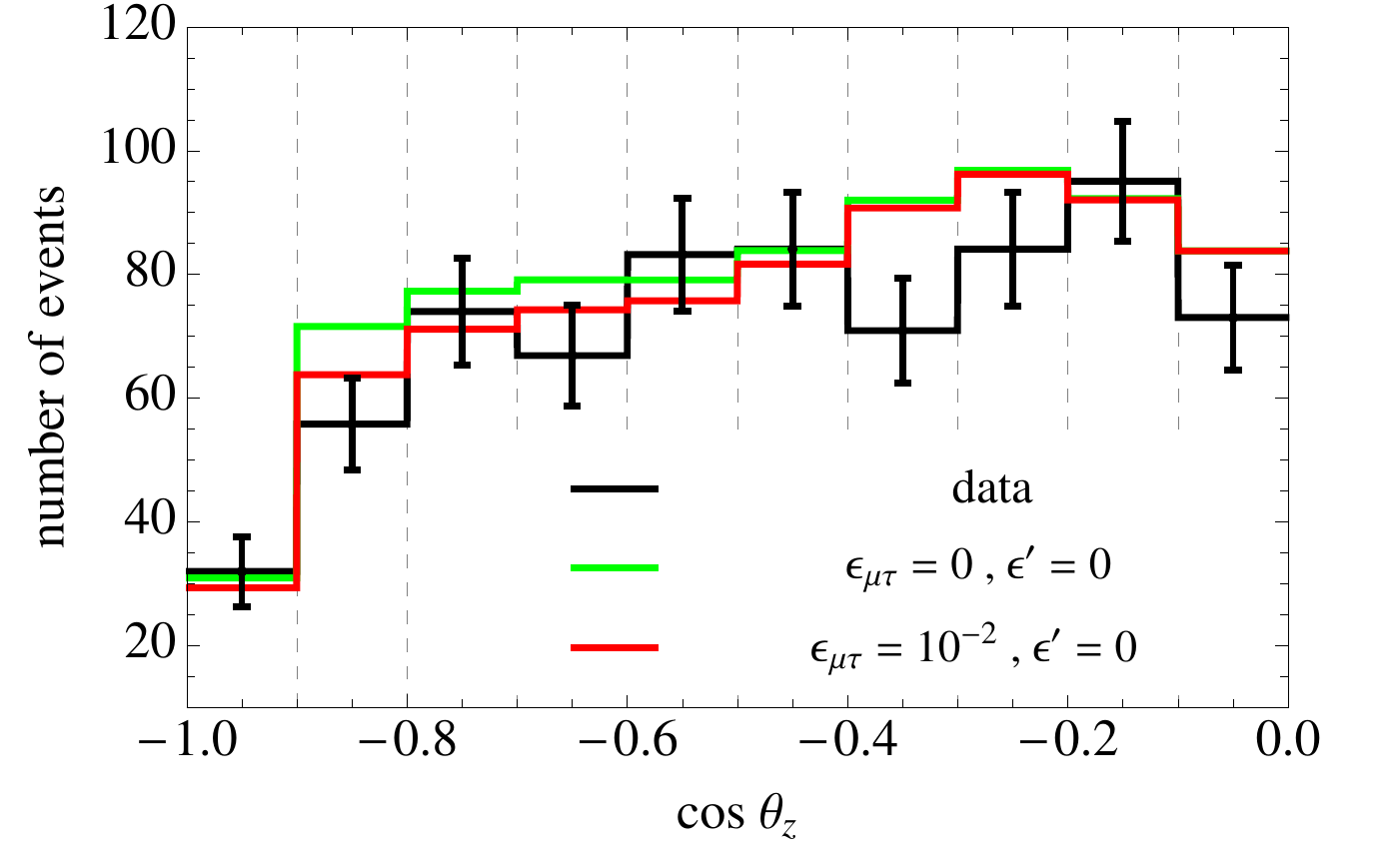}
 \label{fig:low6}
}
\subfloat[]{
 \includegraphics[width=0.5\textwidth]{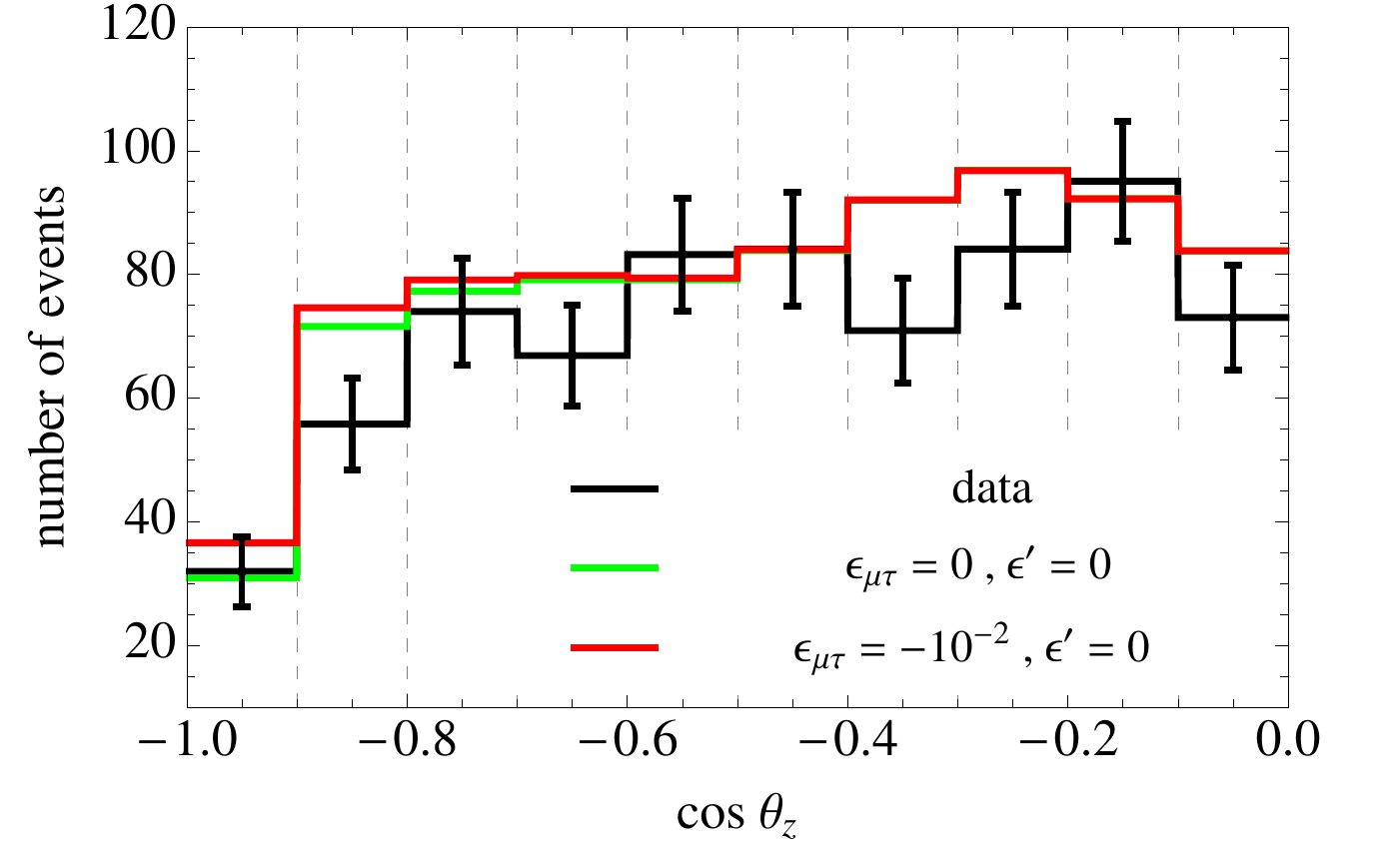}
 \label{fig:low16}
}
\caption{\label{fig:low616}Zenith angle distribution of the low energy $\nu_\mu-$events at DeepCore. Shown are the predicted number of events with NSI characterized by (a) $\epsilon_{\mu\tau}= 0.01$, (b) $\epsilon_{\mu\tau}= -0.01$, and $\epsilon^\prime=0$ (red) and without NSI (green). The black histogram is the DeepCore data with statistical errors.}
\end{figure}

\begin{figure}[t!]  
\centering
\subfloat[]{
 \includegraphics[width=0.5\textwidth]{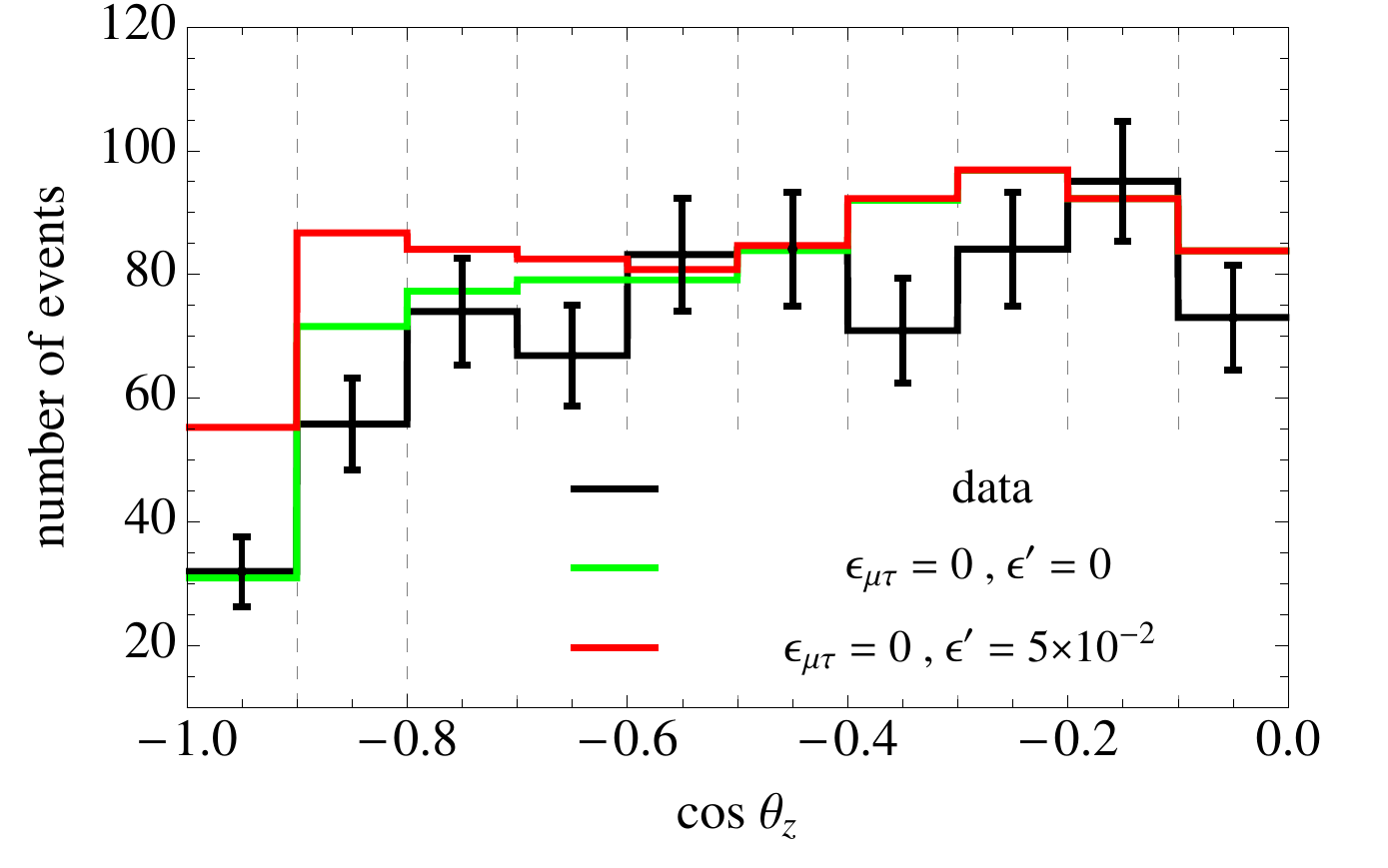}
 \label{fig:low23}
}
\subfloat[]{
 \includegraphics[width=0.5\textwidth]{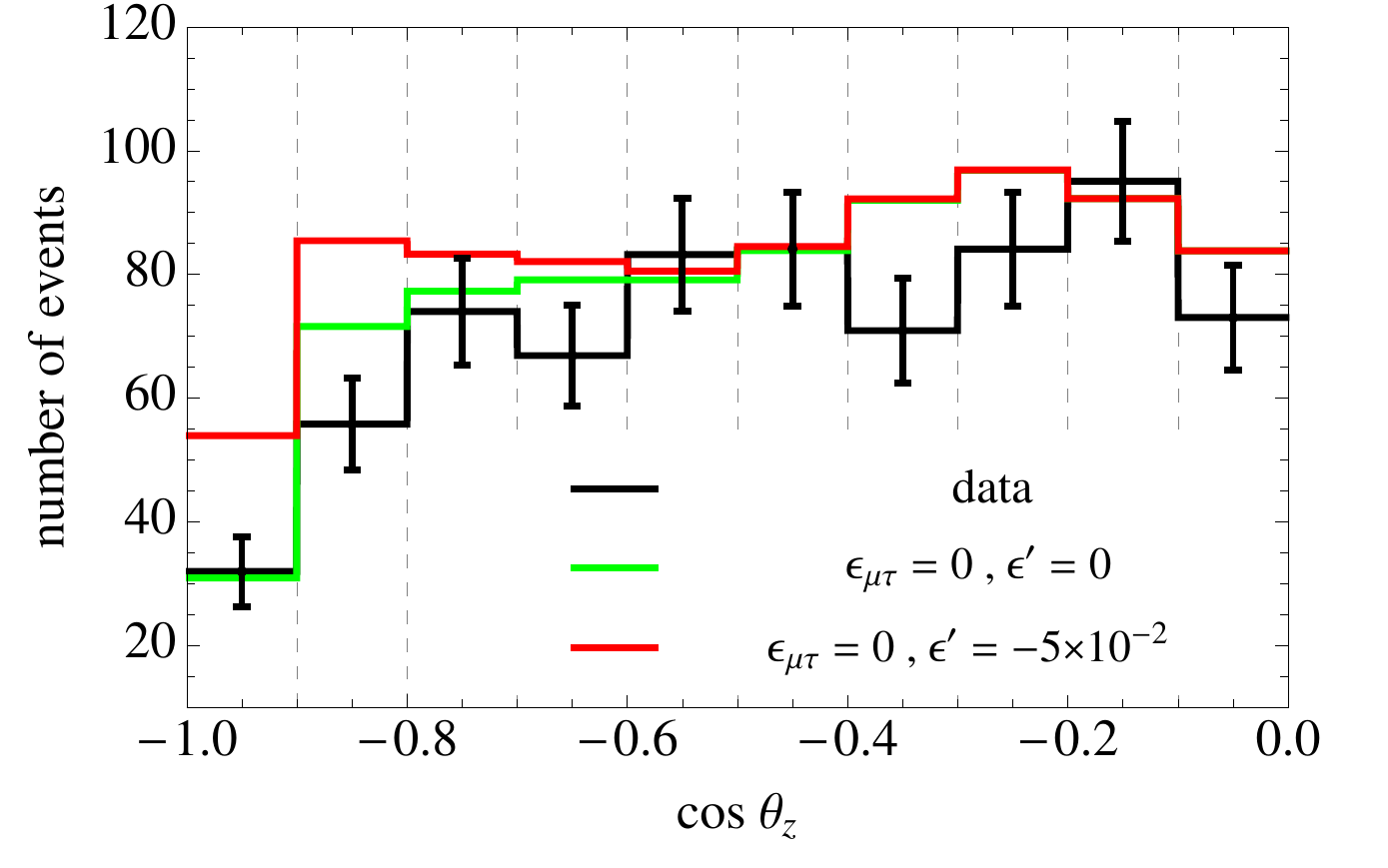}
 \label{fig:low19}
}
\caption{\label{fig:low2324}The same as Fig.~\ref{fig:low616} for NSI characterized by (a) $\epsilon^\prime= 5\times 10^{-2}$ and (b) $\epsilon^\prime= - 5 \times 10^{-2}$ and $\epsilon_{\mu\tau}= 0$.}
\end{figure}

\begin{figure}[t!]
\centering 
\subfloat[]{
 \includegraphics[width=0.5\textwidth]{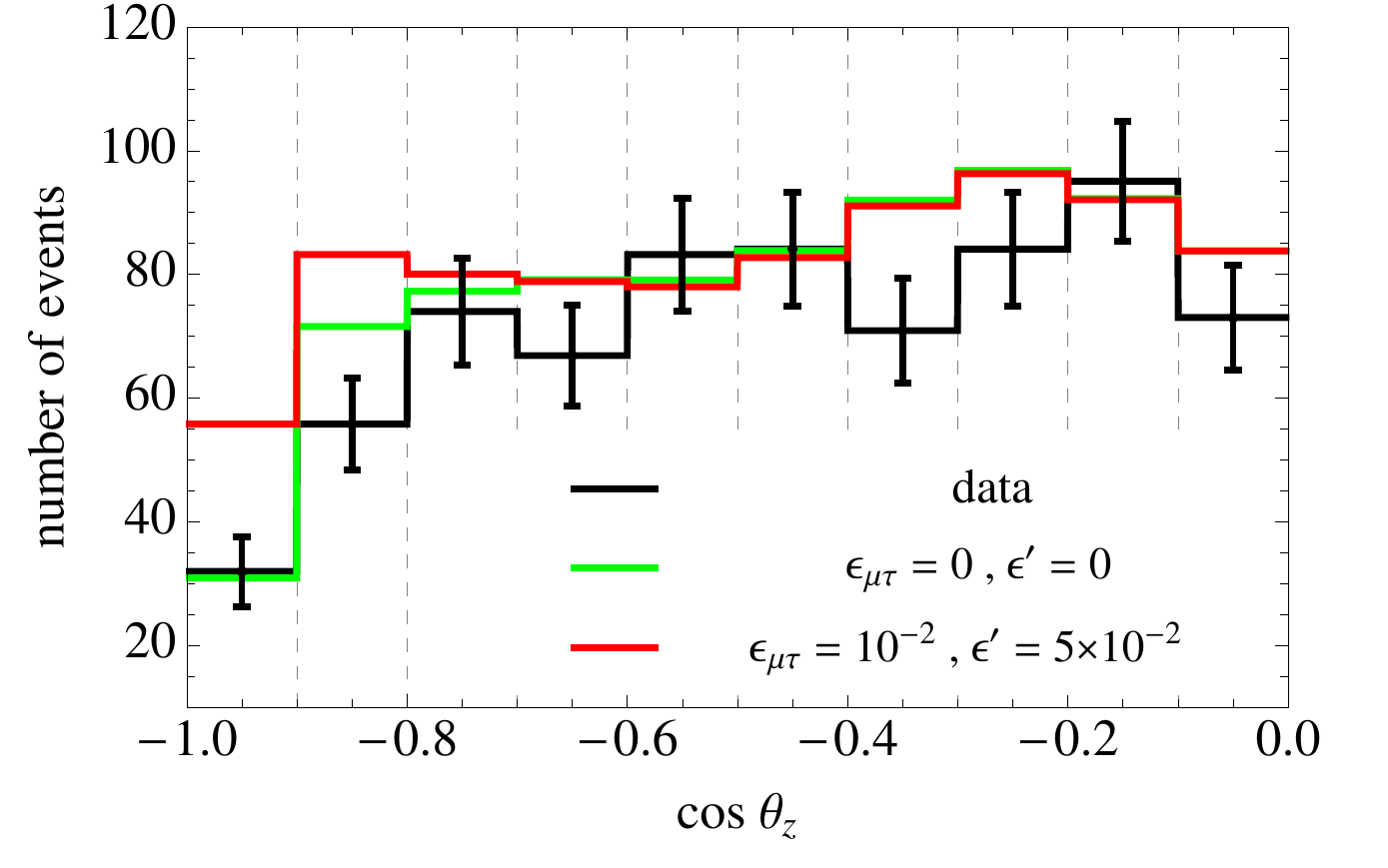}
 \label{fig:low9}
}
\subfloat[]{
 \includegraphics[width=0.5\textwidth]{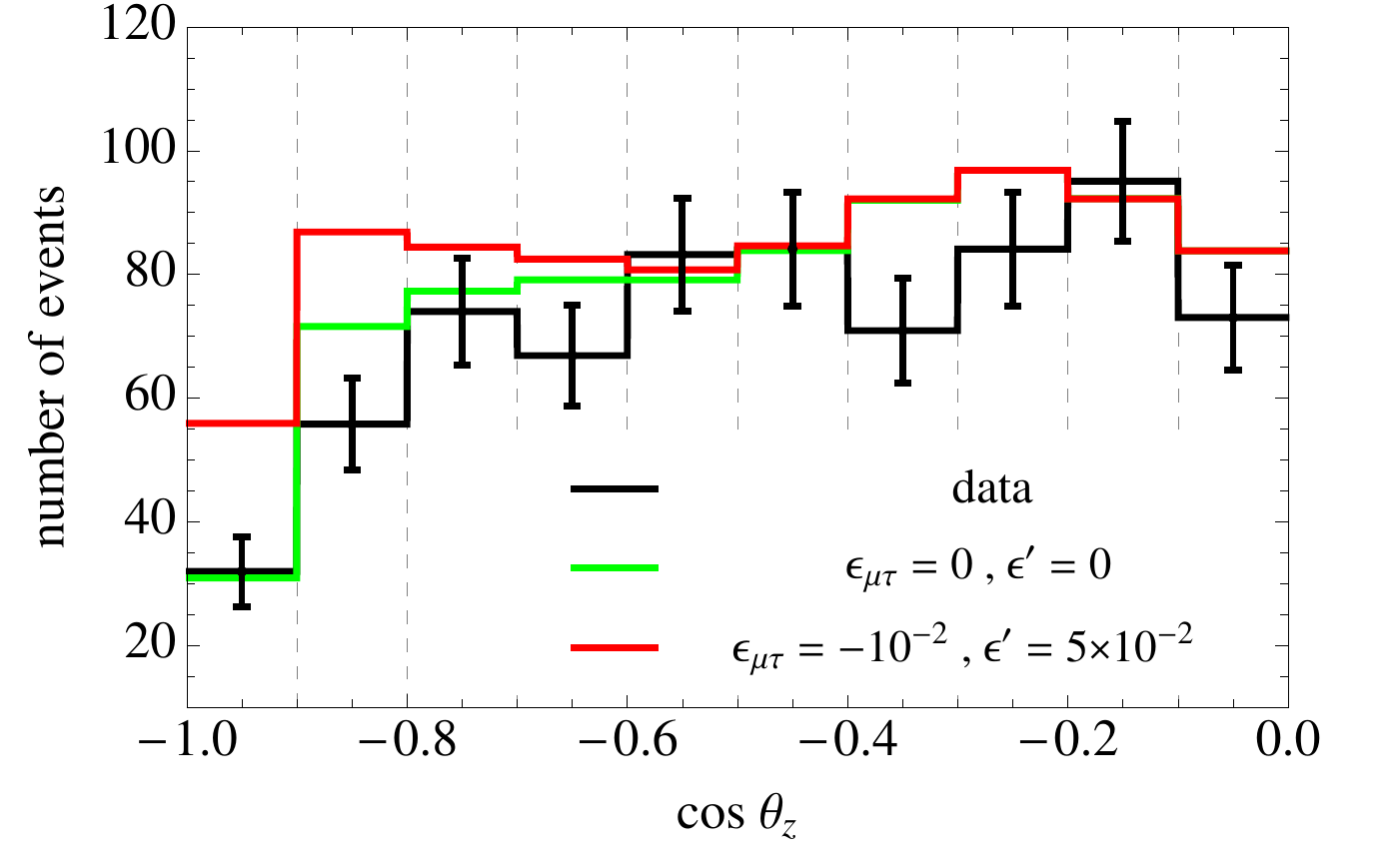}
 \label{fig:low19}
}
\caption{\label{fig:low919}The same as Fig.~\ref{fig:low616} for NSI characterized by (a) $\epsilon_{\mu\tau}=0.01$, (b) $\epsilon_{\mu\tau}= - 0.01$, and $\epsilon^\prime= 5 \times 10^{-2}$.}
\end{figure}

In Figs.~\ref{fig:low616}, \ref{fig:low2324} and \ref{fig:low919} we present zenith angle distributions for the DeepCore (low energy data). In Fig.~\ref{fig:low616} we show effect of $\epsilon_{\mu\tau}$ which is relatively small. For $\epsilon_{\mu\tau} > 0$ (resonance in the $\bar{\nu}$ channel) NSI further suppresses the number of events with the strongest effect in the bin $[-0.9,-0.8]$. This is because NSI shifts the oscillatory pattern to higher energies (see oscillograms in Sec.~\ref{sec:prob}). For $\epsilon_{\mu\tau} < 0$ (resonance in the $\nu$ channel) the oscillatory pattern  shifts to lower energies, effect is smaller and has an opposite sign (number of events increases). Fig.~\ref{fig:low2324} shows effect of $\epsilon^{\prime}$. In this figure NSI reduce the oscillation effect which leads to an increase in the number of events for both signs of $\epsilon^{\prime}$. Indeed, according to our considerations in Sec.~\ref{sec:prob}, the effect of NSI in this case is essentially reduced to suppression of oscillations at high energies relevant for DeepCore. Fig.~\ref{fig:low919} shows the combined effect of $\epsilon_{\mu\tau}$ and $\epsilon^{\prime}$. The result can be understood from the two previous figures, and as can be seen, the effect of $\epsilon^{\prime}$ dominates. 

\subsection{Constraints on $\epsilon_{\mu\tau}$ and $\epsilon^\prime$ from IceCube-79 atmospheric neutrino data}
\label{sec:ic79}

To find bounds on the NSI parameters we define the $\chi^2$ function for the low and high energy intervals as 
\begin{equation}\label{eq:chi2ic40}
\chi^2_{\rm low, high}(\{\epsilon\};\alpha,\beta)  = \sum_i \frac{\left\{N^{\rm 
data}_i-\alpha [1+\beta(0.5+(\cos\theta_z)_i)]  N^{\rm 
exp}_i(\{\epsilon\})\right\}^2}{\sigma_i^2} +  \frac{(1-\alpha)^2}{\sigma_\alpha^2} + 
\frac{\beta^2}{\sigma_\beta^2} , 
\end{equation}
where $\sigma_\alpha=0.25$ is the normalization error of the atmospheric neutrino flux and $\sigma_\beta=0.04$ is the slope error in the zenith angle dependence of the flux~\cite{Honda:2006qj}. $N_i^{\rm data}$ is the observed number of events and $N_i^{\rm exp}(\{\epsilon\})$ is the expected number of events in the presence of NSI.

\begin{figure}[t!]
\centering
\subfloat[]{
 \includegraphics[width=0.5\textwidth]{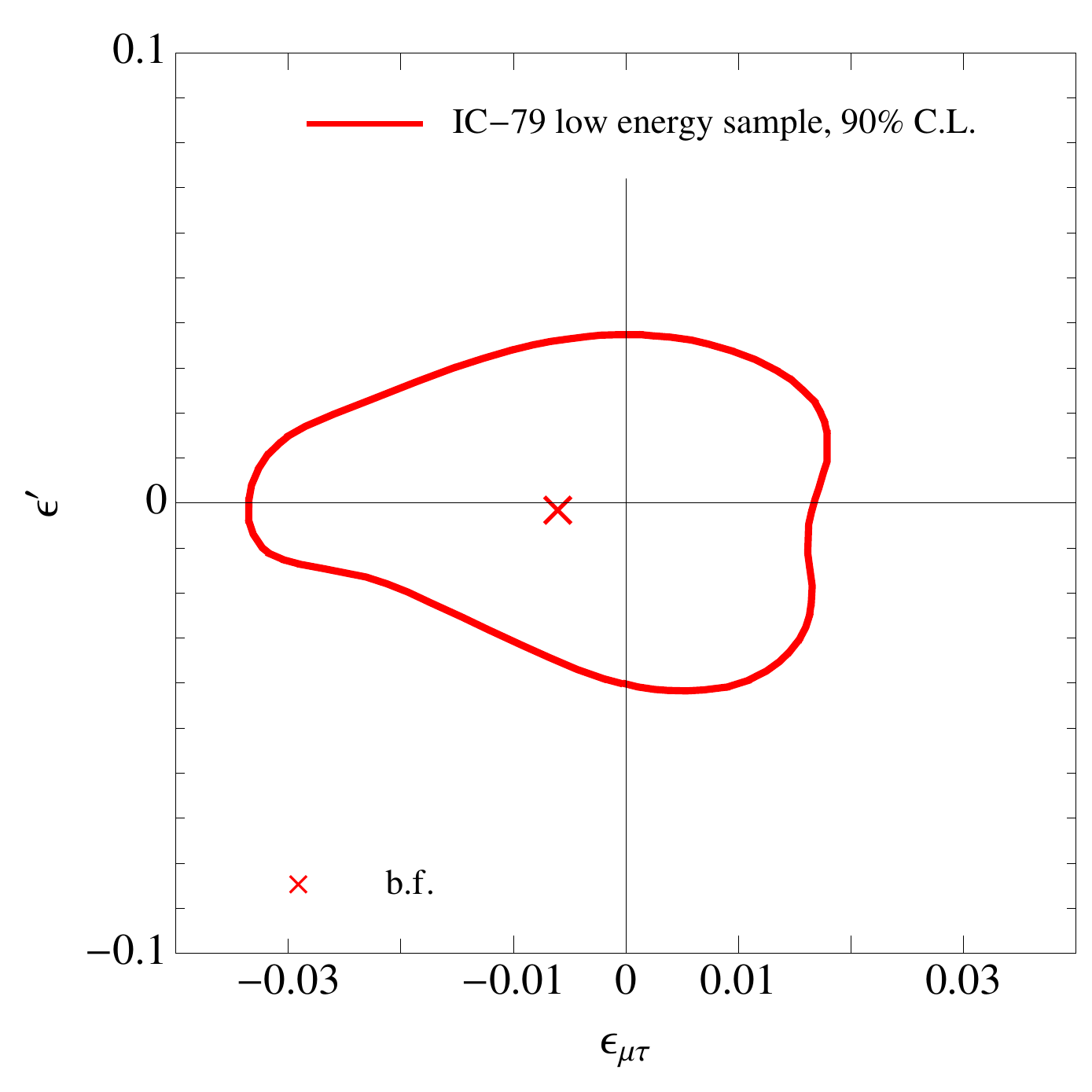}
 \label{fig:ic79low}
}
\subfloat[]{
 \includegraphics[width=0.5\textwidth]{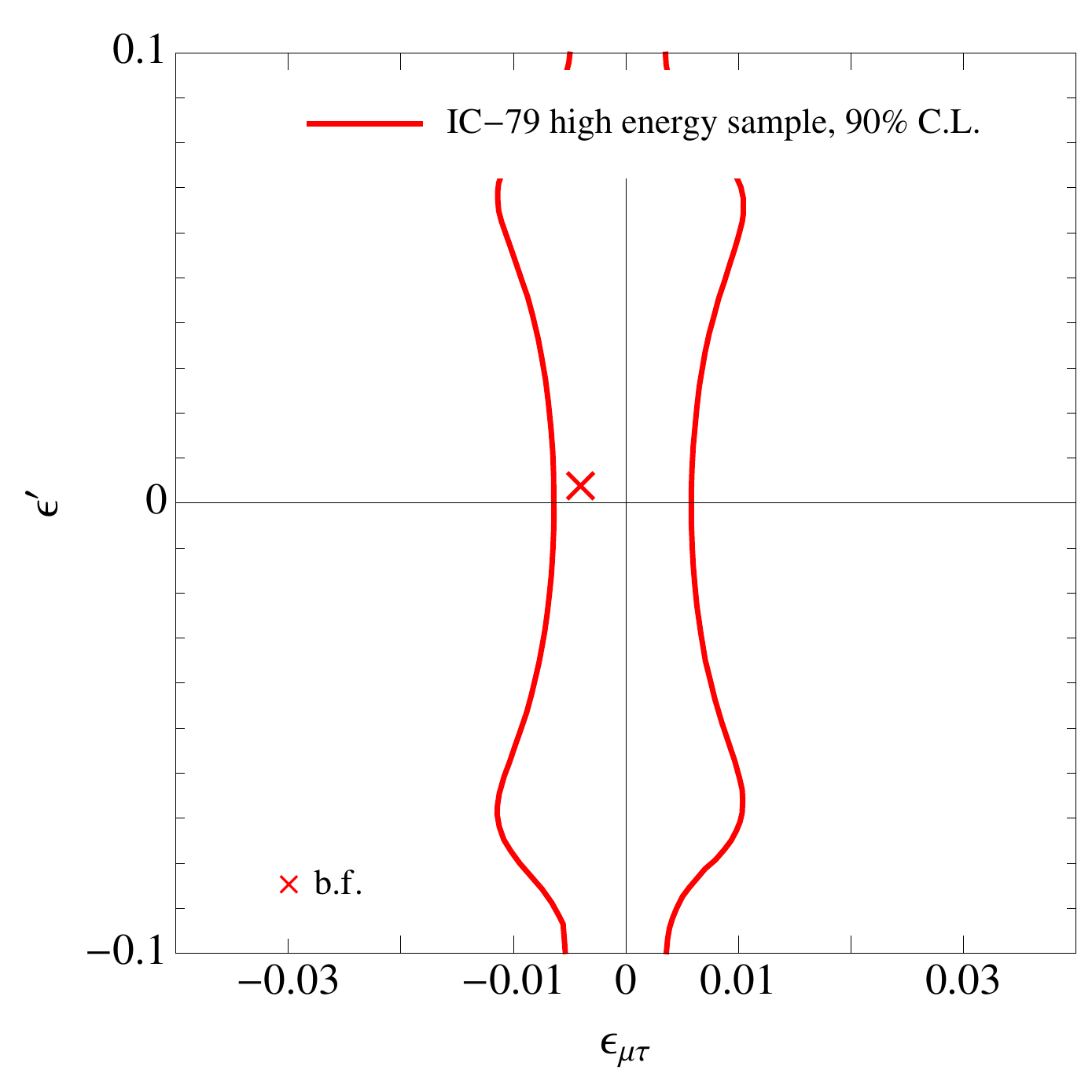}
 \label{fig:ic79high}
}
\caption{\label{fig:ic79lowandhigh}Allowed region in the $(\epsilon_{\mu\tau},\epsilon^\prime)$ plane for the low and high energy data sample of IceCube-79. The cross in each figure shows best-fit value.}
\end{figure}

As expected (see discussion in Sec.~\ref{sec:prob}), the high energy data sample has very good sensitivity to $\epsilon_{\mu\tau}$ but is practically insensitive to $\epsilon^\prime$. The low energy sample of data has comparable sensitivities to $\epsilon^\prime$ and $\epsilon_{\mu\tau}$. These 
features can be seen from the allowed regions of the parameters shown in Fig.~\ref{fig:ic79lowandhigh}. For the low energy and high energy samples, the value of $\Delta \chi^2$ between best-fit and $\{\epsilon\}=0$ is $0.5$ and $2.4$, respectively.

\begin{figure}[t!]
\begin{center}
 \includegraphics[width=0.6\textwidth]{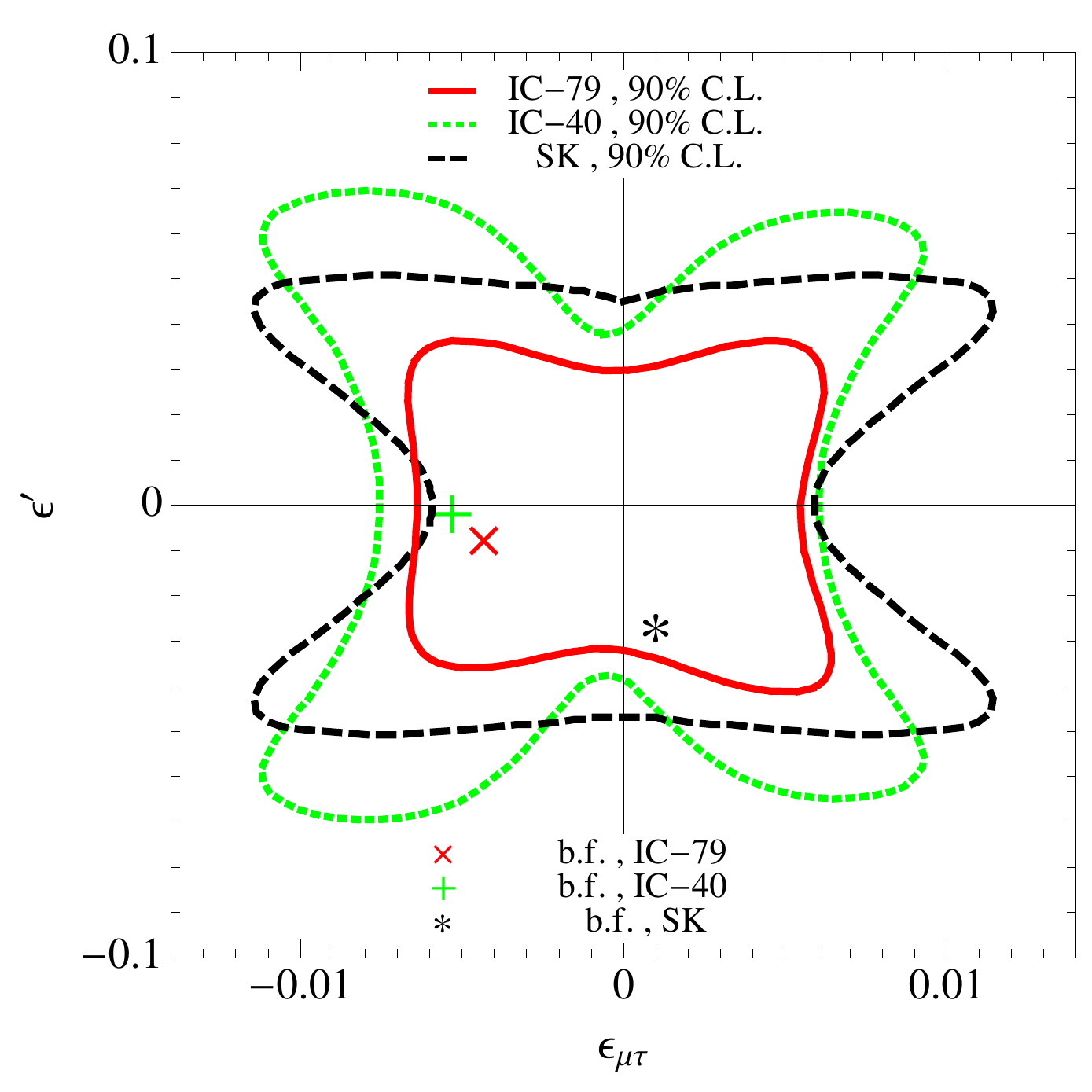}
\end{center}
\caption{\label{fig:allowedic79}
Allowed region in the plane of $(\epsilon_{\mu\tau},\epsilon^\prime)$ at 90\% C.L. obtained from the combined analysis of low and high energy samples of data (IceCube-79 and DeepCore respectively), shown by red solid curve. The black dashed curve shows the allowed region from Super-Kamiokande experiment, taken from~\cite{Mitsuka:2011ty}, and the green dotted curve is for IceCube-40. The red ``$\times$", green ``+" and black ``$\ast$" signs show the best-fit values of NSI parameters from IceCube-79, IceCube-40 and Super-Kamiokande experiments, respectively.}
\end{figure}

Combining the low (DeepCore) and high energy (IceCube-79) analyses, it is possible to constrain both NSI parameters $\epsilon_{\mu\tau}$ and $\epsilon^\prime$ to a greater degree. Fig.~\ref{fig:allowedic79} shows the allowed region obtained from the combined analysis of the low and high energy sample (red solid curve). In this figure we have shown also the allowed region from IceCube-40, which is obtained from the analysis of atmospheric neutrinos in the energy range of $100$~GeV to $400$~TeV~\cite{Abbasi:2010ie}. For the IceCube-40 analysis we used the same $\chi^2$ function as in Eq.~(\ref{eq:chi2ic40}). The black dashed curve in Fig.~\ref{fig:allowedic79} corresponds to the allowed region from Super-Kamiokande experiment~\cite{Mitsuka:2011ty}. As can be seen the limit from IceCube-79 is stronger than the one from IceCube-40 due to higher statistics and inclusion of the lower energy data. We have checked that combining the IceCube-79 and IceCube-40 do not improve the limit significantly.

After marginalizing with respect to $\epsilon^\prime$ ($\epsilon_{\mu\tau}$) we obtain the following allowed ranges for $\epsilon_{\mu\tau}$ and $\epsilon^\prime$ at 90\% confidence level:
\begin{eqnarray}
\label{eq:limitic79}
- 6.1 \times 10^{-3} < \epsilon_{\mu\tau} < 5.6\times10^{-3}, 
~~~~~~ 90\% \;{\rm C.L.} \\ 
~ -3.6\times10^{-2}<\epsilon^\prime < 3.1\times10^{-2},  
~~~~~ 90\% \;{\rm C.L.} .  
\label{eq:limitic791}
\end{eqnarray}

The high statistics data of IceCube-79 ($E_\nu > 100$~GeV) can be used to constrain the normalization uncertainty of atmospheric neutrino flux (represented by $\alpha$ in Eq.~(\ref{eq:chi2ic40})) and so increase the sensitivity of DeepCore data to NSI. We performed such an analysis which leads to stronger bounds on NSI strength parameters than in Eqs.~(\ref{eq:limitic79}) and (\ref{eq:limitic791}). However, since the systematic errors of IceCube detector is not available yet and the normalization uncertainty would be energy dependent, we report in this paper the more conservative limits in Eqs.~(\ref{eq:limitic79}) and (\ref{eq:limitic791}).

According to Fig.~\ref{fig:allowedic79}, in spite of higher energies of events, the IceCube-79 bound is comparable with the Super-Kamiokande bound. The reason is that the IceCube-79 limit has been obtained by analyzing the zenith angle distribution of $\nu_\mu-$events (that is integrated  over energy) only;
while in the Super-Kamiokande analysis both the zenith angle and energy distributions of $\nu_\mu-$events have been taken into account. Thus, we expect that stronger limit from IceCube data than the one in Fig.~\ref{fig:allowedic79} can be obtained by performing the energy analysis of $\nu_\mu-$events.

The limits in Eqs.~(\ref{eq:limitic79}) and (\ref{eq:limitic791}) are the strongest available limits on the NSI parameters $\epsilon_{\mu\tau}$ and $\epsilon^\prime$. Specifically, the limit on $\epsilon_{\mu\tau}$ in Eq.~(\ref{eq:limitic79}) is stronger than the Super-Kamiokande limit by a factor of two.  

We have performed an analysis of the data assuming $15\%$ uncorrelated systematic errors in each zenith angle bin. This weakens the bounds on $\epsilon_{\mu\tau}$ and $\epsilon^\prime$ by factor 1.6.

\subsection{Future sensitivity of DeepCore to $\epsilon_{\mu\tau}$ and $\epsilon^\prime$}
\label{sec:deepcore}
 
Future operation of IceCube and especially DeepCore will improve the sensitivity to the NSI parameters $\epsilon_{\mu\tau}$ and $\epsilon^\prime$. This improvement will be achieved due to higher statistics, lower energy threshold of event selection and a possibility to explore the energy dependence of the zenith angle distribution below $100$~GeV. 

In this section we calculate the sensitivity of DeepCore to the NSI parameters taking $E_\nu=10$~GeV as the energy threshold. We assume a factor of two uncertainty for the neutrino energy resolution (accuracy of reconstruction), that could be reflected by selection of the corresponding energy bins: we take three bins: $[10-20]$~GeV, $[20-40]$~GeV and $[40-80]$~GeV. IceCube has high precision in the reconstruction of incoming $\nu_\mu$ ($\bar{\nu}_\mu$) direction ($\sim 1^\circ$). However, in the low energy range the reconstruction of 
incoming neutrino direction is limited by the uncertainty due to the scattering angle of neutrinos off nuclei as well as worser reconstruction of the muon direction. It can be approximated by $\Delta 
\theta_z=1.1\sqrt{m_p/(E_\nu/{\rm GeV})}$, so that $\Delta \theta_z$ varies from $19^\circ$ at $E_\nu=10$~GeV to $7^\circ$ at $E_\nu=80$~GeV. For simplicity we assume the worst resolution (that is $19^\circ$) for the entire energy range $(10-80)$~GeV. Correspondingly, we consider four bins of $\cos\theta_z$: $[-1,-0.9]$, $[-0.9,-0.6]$, $[-0.6,-0.3]$ and $[-0.3,0]$. The number of events in the $i$-th bin of $\cos\theta_z$ and $j$-th bin of $E_\nu$ is given by:\\
 
\begin{figure}[t!]
\begin{center}
 \includegraphics[width=0.5\textwidth]{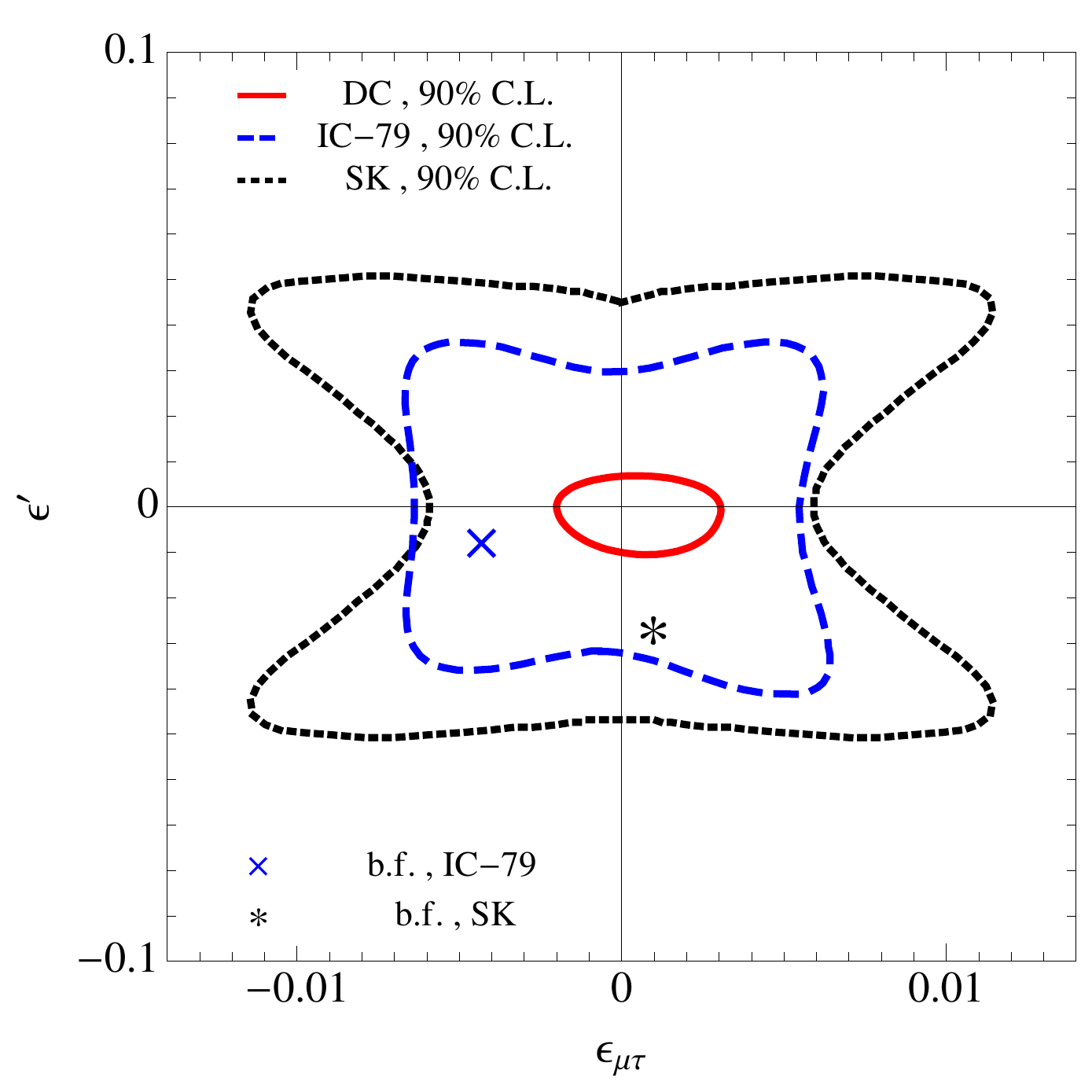}
\end{center}
\caption{\label{fig:alloweddc}Sensitivity of DeepCore in the plane of $(\epsilon_{\mu\tau},\epsilon^\prime)$ at 90\% C.L. after one year of data-taking (shown by red solid curve). The black dotted and blue dashed curves show the allowed region from Super-Kamiokande and IceCube-79+DeepCore experiments, respectively. The blue ``$\times$" and black ``$\ast$" signs show the best-fit values of NSI parameters from IceCube-79 and Super-Kamiokande experiments, respectively.}
\end{figure}

\begin{figure}[t!]
\centering
\subfloat[]{
 \includegraphics[width=0.5\textwidth]{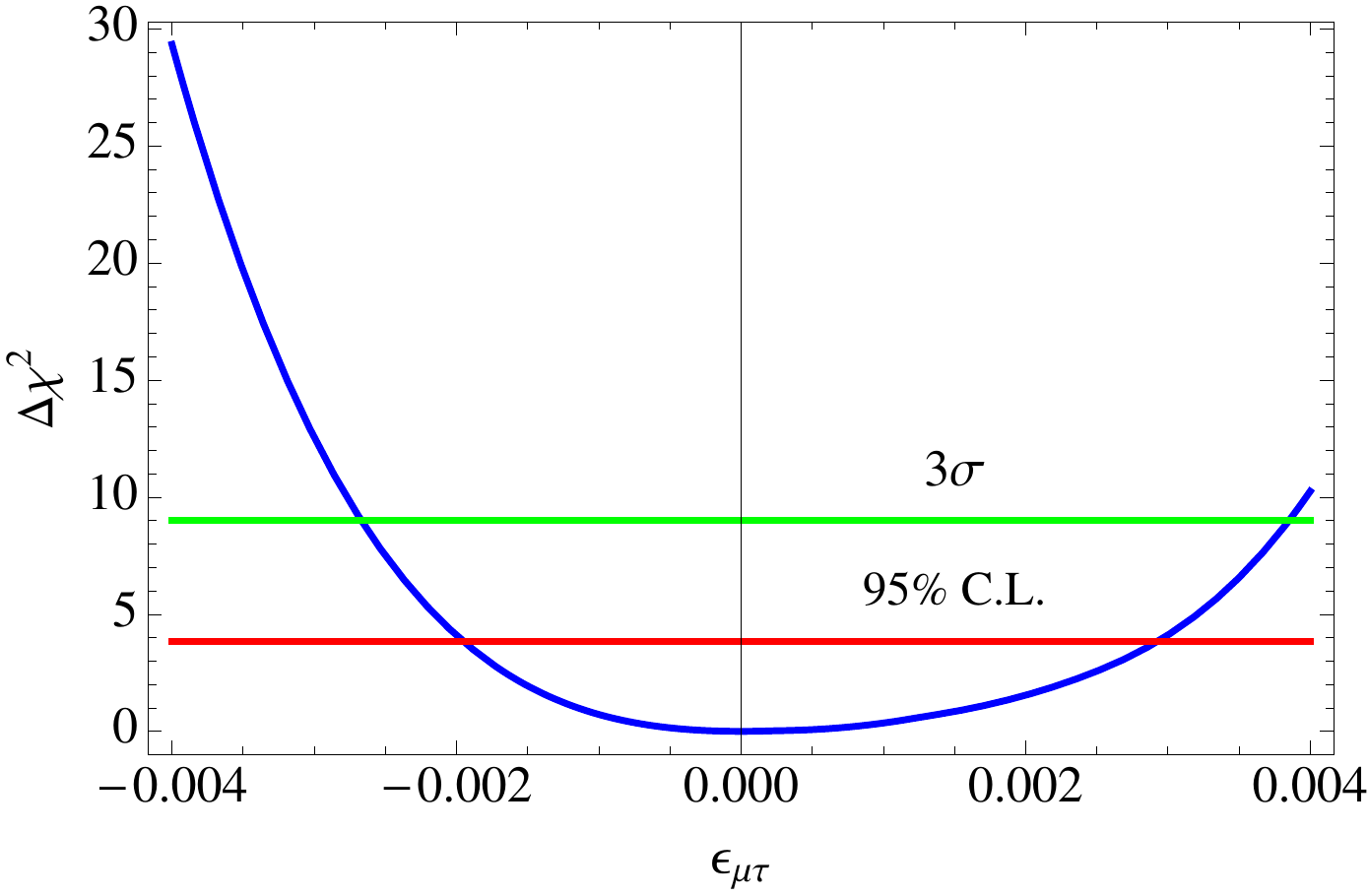}
 \label{fig:chi2dca}
}
\subfloat[]{
 \includegraphics[width=0.5\textwidth]{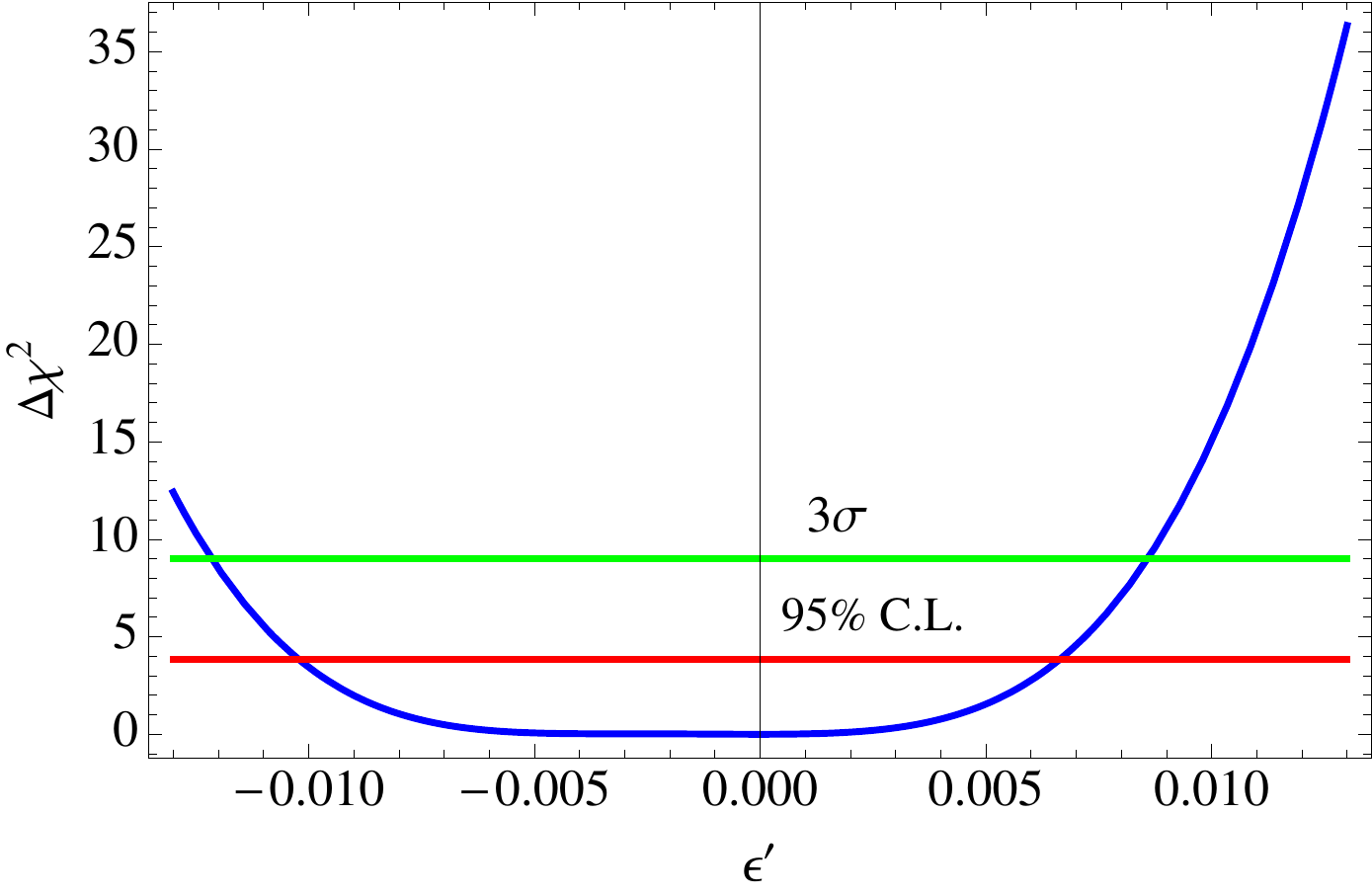}
 \label{fig:chi2dcb}
}
\caption{\label{fig:chi2dc}$\Delta \chi^2$, from Eq.~(\ref{eq:chi2dc}), as a function of $\epsilon_{\mu\tau}$ and $\epsilon^\prime$, for DeepCore experiment after one year of data-taking.}
\end{figure}

\begin{eqnarray}
&  & \!\!\!\!\!\!\!\!\!\!\!\!\!\!\! N_{i,j}(\{\epsilon\}) = T\Delta\Omega \int_{\Delta_i \cos\theta_z} d\cos\theta_z \int_{\Delta_j E_\nu}  dE_\nu \\
&  & \left[ \Phi_{\nu_\mu}(E_\nu,\cos\theta_z) P(\nu_\mu \rightarrow \nu_\mu) (\{\epsilon\}) 
+ \Phi_{\nu_e}  (E_\nu,\cos\theta_z) P(\nu_e \rightarrow  \nu_\mu) (\{\epsilon\}) \right] A_{\rm eff}^{\nu_\mu} (E_\nu,\cos\theta_z) 
\nonumber\\ 
& & + (\nu \to \bar{\nu} )~, \nonumber
\end{eqnarray} 
where for the effective area of DeepCore we use the simulation of~\cite{Collaboration:2011ym} and we assume that the effective area is independent of the zenith angle: $A_{\rm eff}(E_\nu,\theta_z)\equiv A_{\rm eff}(E_\nu)$. Also we assume the same effective area for muon neutrino and antineutrino: $A_{\rm eff}^{\nu_\mu}=A_{\rm eff}^{\bar{\nu}_\mu}$. To calculate the sensitivity of 
DeepCore to the NSI parameters, we define the following $\chi^2$ function:
\begin{equation}\label{eq:chi2dc}
\chi^2_{\rm DC}(\{\epsilon\};\alpha,\beta)= 
\sum_{i,j} \frac{\left\{N_{i,j}(\{\epsilon\}=0)- 
\alpha 
[1+\beta(0.5+(\cos\theta_z)_i)] 
N_{i,j}(\{\epsilon\})\right\}^2}{N_{i,j}(\{\epsilon\})} 
+ \frac{(1-\alpha)^2}{\sigma_\alpha^2}+\frac{\beta^2}{\sigma_\beta^2}, 
\end{equation}
where $\alpha$ and $\beta$ have the same meaning as in Eq.~(\ref{eq:chi2ic40}) (with 
$\sigma_\alpha=0.25$ and $\sigma_\beta=0.04$). With this $\chi^2$ function, after marginalizing with respect to $\alpha$ and $\beta$, we obtain the sensitivity to $\epsilon_{\mu\tau}$ and $\epsilon^\prime$ shown in Fig.~\ref{fig:alloweddc}. As can be seen, after one year of data-taking 
(which is already passed) the DeepCore experiment can exclude a large part of the allowed regions from Super-Kamiokande and IceCube-79 experiments.

The marginalized value of $\Delta\chi^2$ with respect to $\epsilon^\prime$ ($\epsilon_{\mu\tau}$), as a function of $\epsilon_{\mu\tau}$ ($\epsilon^\prime$) is shown in Figs.~\ref{fig:chi2dc}. The sensitivity of DeepCore experiment is then
\begin{equation}
-1.7\times10^{-3} < \epsilon_{\mu\tau} <  2.6\times10^{-3} \;  ~~~~~\; 9.5\times10^{-3} < \epsilon^\prime <  6.\times10^{-3}  \quad  90\% \;{\rm C.L.} 
\end{equation}
\begin{equation}
-2.7\times10^{-3} < \epsilon_{\mu\tau} < 3.9\times10^{-3} \; 
~~~~~ \;  -1.2\times10^{-2} < \epsilon^\prime <  8.6\times10^{-3}  \quad  3 \sigma ~{\rm C.L.} .  
\end{equation}
Comparing these result with those in Eqs.~(\ref{eq:limitic79}) and (\ref{eq:limitic791}) we find that sensitivity can be improved by factor $2 - 3$ with one year data of DeepCore.

\section{Discriminating NSI from sterile neutrino effects}
\label{sec:discriminate}

Apart from NSI, new oscillation effects at high energies can be induced by mixing of $\nu_\mu$ with sterile neutrinos, $\nu_s$, composed mainly of the mass state with $m =\mathcal{O}(1)$~eV. The sterile neutrinos are motivated by various anomalies observed in neutrino experiments~\cite{Aguilar:2001ty,AguilarArevalo:2012va,Giunti:2010zu,Mention:2011rk} (see~\cite{Kopp:2013vaa} for a recent global analysis). The oscillations of the active to sterile neutrinos lead to distortion of zenith angle distribution of the $\nu_\mu-$events~\cite{sterile,Esmaili:2012nz} which can be quite similar to distortion due to NSI. Effects of $\nu_s$ have, however, certain features which allow 
to disentangle them from NSI effects. In particular, the energy dependence of the distortions is 
different in the two cases.  

To illustrate this we consider the $3+1$ scenario with $\Delta m_{41}^2=1~{\rm eV}^2$ and the mixing between the sterile and active neutrinos $\sin^2 2\theta_{\mu\mu}\equiv4|U_{\mu4}|^2(1-|U_{\mu4}|^2)=0.1$. The existence of a sterile neutrino with these parameters leads to the MSW active-sterile conversion of the nearly up-going $\nu_\mu$ with energy $\sim3$~TeV~\cite{sterile} (the resonance happens for $\nu_\mu$ if $\Delta m_{41}^2<0$ and for $\bar{\nu}_\mu$ if $\Delta m_{41}^2>0$). Thus, assuming $\Delta m_{41}^2>0$, the resonant conversion $\bar{\nu}_\mu\to\bar{\nu}_s$ results in a reduction of the $\bar{\nu}_\mu$ atmospheric flux at $\cos\theta_z\sim -1$. The oscillograms representing the survival probability of $\bar{\nu}_\mu$ and $\nu_\mu$ can be found in~\cite{Esmaili:2012nz}.

The DeepCore and IceCube allow to measure the zenith angle distributions of events in different energy intervals. In this connection we calculate the zenith distribution of the $\nu_\mu-$events in DeepCore and IceCube for two cases: 
1) NSI;  
2) the $3+1$ model.  
We assume a factor of two uncertainty in the reconstruction of neutrino energy, so the energy bins have the width $[E_\nu,2E_\nu]$. For the incoming direction of neutrinos we take $\Delta \cos\theta_z=0.1$, which is realistic for $E_\nu\gtrsim100$~GeV. Since we are performing here an illustrative analysis to clarify the special signatures of the NSI and $3+1$ model, we assume the same resolution of direction reconstruction for the whole energy range in consideration.  

\begin{figure}[t!]
\centering
\subfloat[]{
 \includegraphics[width=0.5\textwidth]{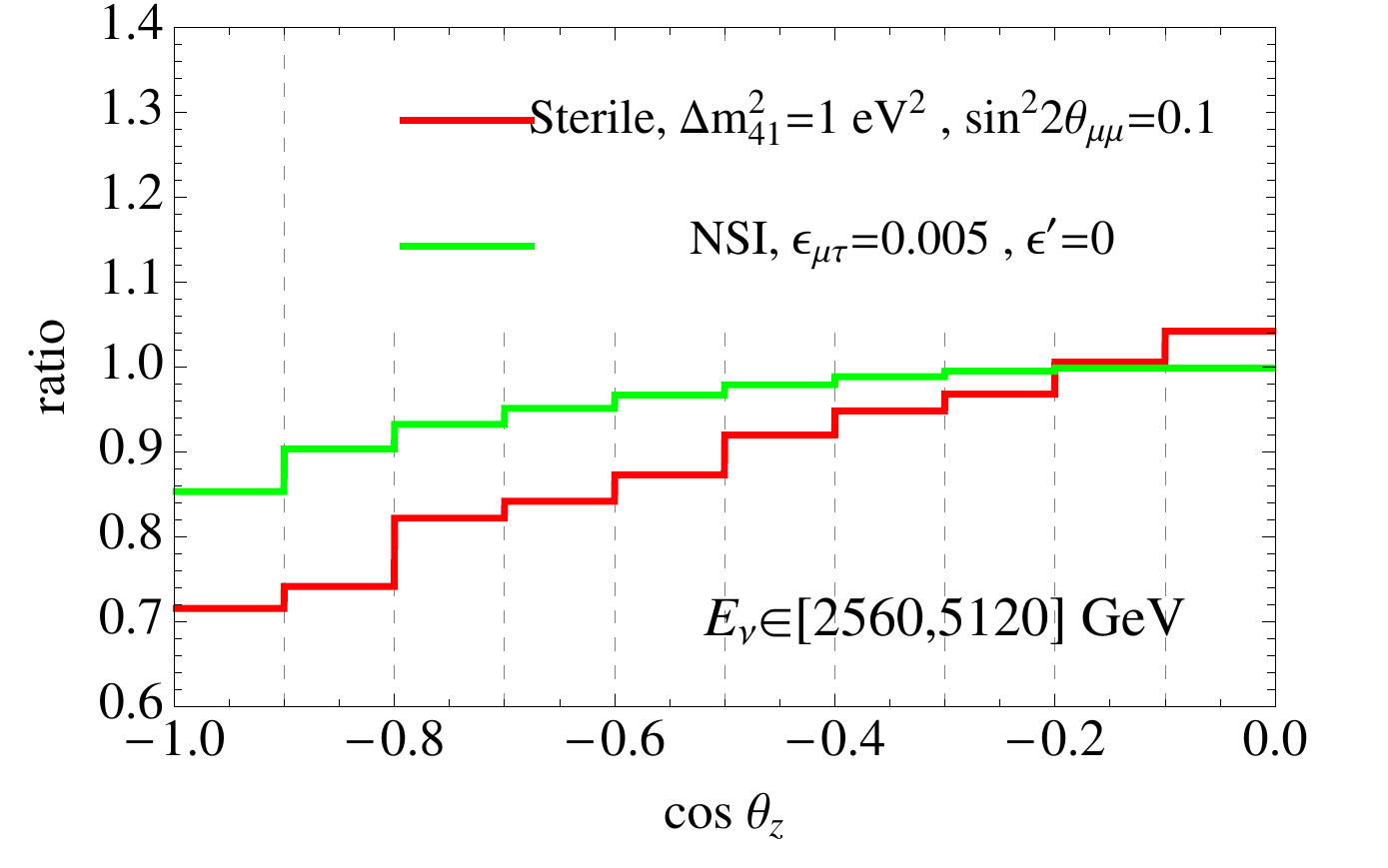}
 \label{fig:ratio90comf}
}
\subfloat[]{
 \includegraphics[width=0.5\textwidth]{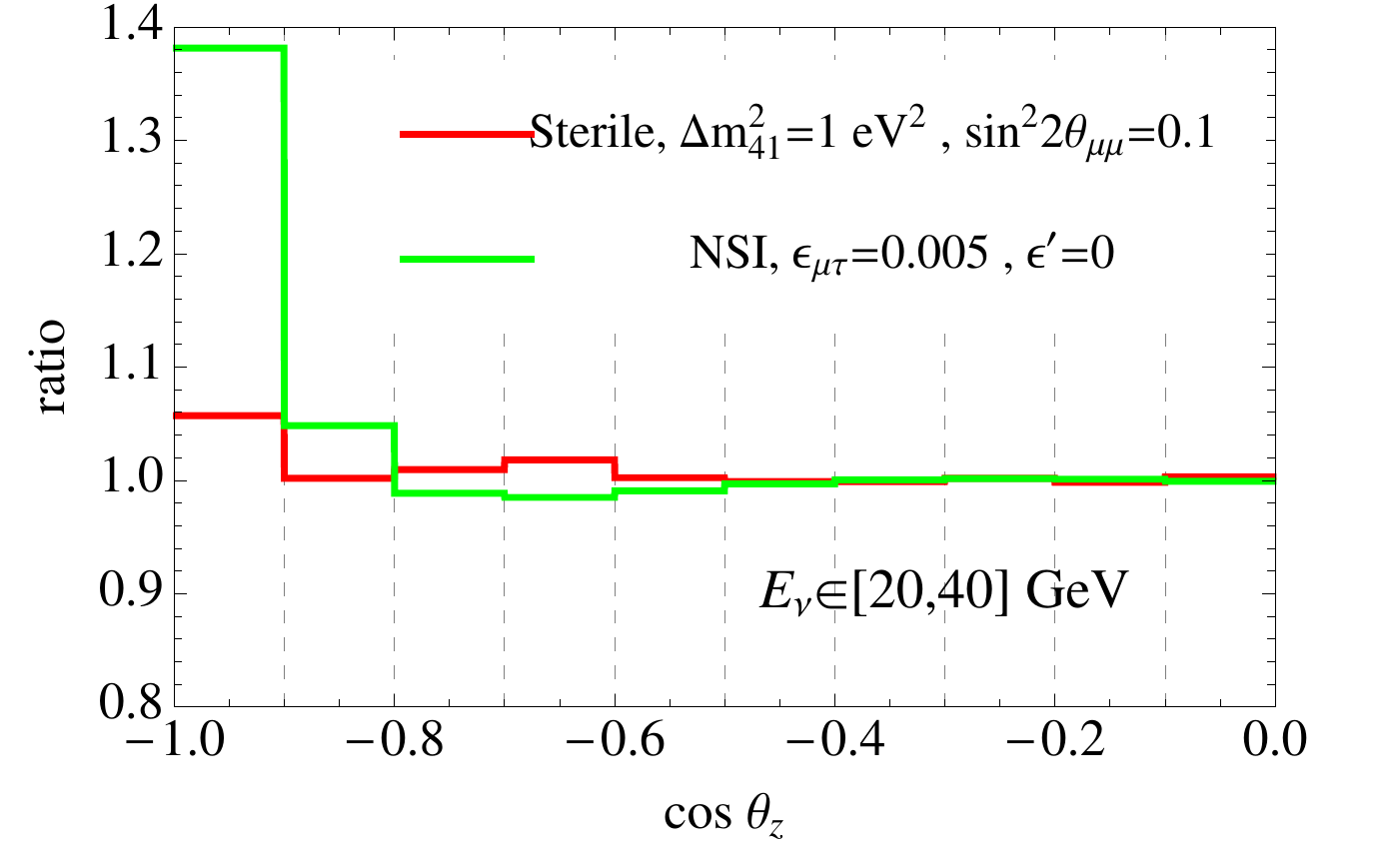}
 \label{fig:ratio90comb}
}
\caption{\label{fig:ratiocompare901}The ratios of the numbers of events for NSI and $3+1$ model for the IceCube (a) and DeepCore (b). We take for $3+1$ model $\Delta m_{41}^2=1.0~{\rm eV}^2$ and $\sin^22\theta_{\mu\mu}=0.1$, and for NSI: $\epsilon_{\mu\tau}=5\times10^{-3}$ and $\epsilon^\prime=0$.}
\end{figure}

Fig.~\ref{fig:ratiocompare901} presents the distortions of the zenith angle distributions for the two different bins of neutrino energy. We show the ratios $N(\{\epsilon\})/N^{\rm STD}_{3\nu}$ for the NSI parameters $\epsilon_{\mu\tau}=5\times10^{-3}$ and $\epsilon^\prime = 0$; and $N(\Delta m_{41}^2,\sin^22\theta_{\mu\mu})/N^{\rm STD}_{3\nu}$ for the $3+1$ model parameters $\Delta m_{41}^2=1.0~{\rm eV}^2$ and $\sin^22\theta_{\mu \mu} = 0.1$. Here $N^{\rm STD}_{3\nu}$ is the number of events in the standard $3\nu$ framework. 

There are two salient features which distinguish NSI from sterile neutrinos. 
\begin{enumerate}
\item For the assumed values of $3+1$ parameters, the MSW resonance occurs at $E_\nu\sim3$~TeV, which lies within the high energy bin of Fig.~\ref{fig:ratio90comf}. Oscillations to sterile  neutrinos lead to strong suppression of signal as compared with NSI effect. There is no such a strong effect in other energy bins. 

\item Another feature is the peak in zenith distribution in the bin $\cos\theta_z\in[-1,-0.9]$ and $[20,40]$~GeV, for NSI (see Fig.~\ref{fig:ratio90comb}). The peak can be interpreted in the following way: the bin $[20,40]$~GeV contains the minimum of the survival probability for muon 
(anti)neutrino passing the diameter of Earth. The NSI leads to a shift in the position of minimum to lower (higher) energies for muon (anti)neutrinos, which can be seen in Fig.~\ref{fig:NSIwhole}. In the energy bin $[20,40]$~GeV, in both neutrino and antineutrino channels the probability of  oscillations in the presence of NSI is larger. But, this is not the case for the other bins. 

\end{enumerate}

\section{Conclusion}
\label{sec:conc}

\noindent
1. We studied effects of NSI on oscillations of the high energy ($E_\nu > 20$~GeV) atmospheric neutrinos. The reason for that is that in general, the NSI effects do not disappear with increase of energy up to rather high energies. We focussed mainly on the $\nu_\mu - \nu_\tau$ sector.\\

\noindent
2. Oscillograms for the $\nu_\mu$ transitions have been constructed for different NSI scenarios (values of the NSI strength parameters). NSI's modify the pattern of the oscillograms which depend substantially on the presence of the flavor changing parameters and the sign of parameters. NSI's dominate above the resonance energy: $E_\nu \gg E_R(\epsilon)$. Their effect can be enhanced 
at $E_\nu \sim E_R(\epsilon)$ due to interference with usual oscillations. For low energies NSI produce corrections to the standard oscillations of the order $\epsilon$. At high energies the effect 
disappears as $\epsilon_{\mu \tau}^2$ with decrease of the NSI strength.\\

\noindent
3. NSI lead to distortion of the zenith angle distributions of events, which depends on the neutrino energy at and below $E_R(\epsilon)$. We computed the zenith distributions for the low and high energy samples and confronted them with results from IceCube-79 and DeepCore. The data from IceCube-79 and DeepCore are in agreement with standard oscillations. This allowed us to put new and the most stringent limits on the strength parameters: 
$$
- 6.1 \times 10^{-3} < \epsilon_{\mu\tau} < 5.6\times10^{-3}, 
~~~~~~ 90\% \;{\rm C.L.} $$
$$ 
~ -3.6\times10^{-2}<\epsilon_{\tau\tau}-\epsilon_{\mu\mu} <  3.1\times10^{-2},  ~~~~~ 
90\% \;{\rm C.L.}.  
$$

\noindent
4. Future measurements at DeepCore and also at its upgrades will further improve the bounds. In particular, we showed that measurements of the energy and zenith angle distributions in DeepCore  will allow to strengthen limit by factor $2-3$. 

The effects of NSI in the PINGU detector have been studied in~\cite{Ohlsson:2013epa}. For $\epsilon \sim 0.1$ considered in~\cite{Ohlsson:2013epa} the region of strong NSI effect is at $10$~GeV and so PINGU indeed could have good sensitivity. However for $\epsilon \sim 0.01$ the NSI effect shifts to $20 - 100$~GeV, that is beyond the region considered in~\cite{Ohlsson:2013epa}. \\

\noindent
5. We have shown how effects of NSI can be disentangled from effects of sterile neutrinos with mass in the eV range. The two effects have different energy dependencies. There are two salient signatures: the MSW resonance dip in TeV range for sterile neutrinos, and sharp peak in the zenith angle distribution for NSI in the energy range of the first oscillation minimum due to the 2-3 mass splitting.

\section*{Acknowledgment}

The authors thank CENAPAD and CCJDR for computing facilities. A.~E. thanks FAPESP for financial support.


 
\end{document}